\documentclass[a4paper,fleqn]{cas-dc}

\usepackage[authoryear]{natbib}

\def\tsc#1{\csdef{#1}{\textsc{\lowercase{#1}}\xspace}}
\tsc{WGM}
\tsc{QE}

\begin{document}
\let\printorcid\relax
\let\WriteBookmarks\relax
\def\floatpagepagefraction{1}
\def\textpagefraction{.001}

\shorttitle{}    

\shortauthors{H. Liu et al.}  

\title [mode = title]{UniSurgSAM: A Unified Promptable Model for Reliable Surgical Video Segmentation}  

\author[1]{Haofeng Liu}
\credit{Conceptualization, Methodology, Software, Investigation, Data curation, Writing – original draft \& editing}
\ead{haofeng.liu@u.nus.edu}

\author[2]{Ziyue Wang}
\credit{Conceptualization, Validation, Writing – review \& editing}

\author[1]{Alex~Y.~W. Kong}
\credit{Validation, Software}

\author[1]{Guanyi Qin}
\credit{Writing – review \& editing}

\author[1]{Yunqiu Xu}
\credit{Writing – review \& editing}

\author[1]{Chang~Han Low}
\credit{Data curation, Writing – review}

\author[3]{Mingqi Gao}
\credit{Formal analysis, Writing – review}

\author[4]{Lap~Yan~Lennon Chan}
\credit{Data curation}

\author[1,2]{Yueming Jin}
\corref{cor1}
\cortext[cor1]{Corresponding author.}
\ead{ymjin@nus.edu.sg}
\credit{Supervision, Project administration, Conceptualization, Funding acquisition, Writing – review \& editing}

\affiliation[1]{
    organization={Department of Biomedical Engineering, National University of Singapore},
    city={Singapore},
    country={Singapore}
}

\affiliation[2]{
    organization={Department of Electrical and Computer Engineering, National University of Singapore},
    city={Singapore},
    country={Singapore}
}

\affiliation[3]{
    organization={School of Computer Science, The University of Sheffield},
    city={Sheffield},
    country={UK}
}

\affiliation[4]{
    organization={Department of Computer Science and Engineering, The Chinese University of Hong Kong},
    city={Hong Kong},
    country={China}
}

\nonumnote{This work was supported by the Ministry of Education, Singapore, under the Tier 2 grant (T2EP20224-0028) and the Tier 1 grant (23-0651-P0001).}

\begin{abstract}
Surgical video segmentation is fundamental to computer-assisted surgery. 
In practice, surgeons need to dynamically specify targets throughout extended procedures, using heterogeneous cues such as visual selections, textual expressions, or audio instructions.
However, existing Promptable Video Object Segmentation (PVOS) methods are typically restricted to a single prompt modality and rely on coupled frameworks that cause optimization interference between target initialization and tracking. 
Moreover, these methods produce hallucinated predictions when the target is absent and suffer from accumulated mask drift without failure recovery.
To address these challenges, we present UniSurgSAM, a unified PVOS model enabling reliable surgical video segmentation through visual, textual, or audio prompts. 
Specifically, UniSurgSAM employs a decoupled two-stage framework that independently optimizes initialization and tracking to resolve the optimization interference.
Within this framework, we introduce three key designs for reliability: presence-aware decoding that models target absence to suppress hallucinations; boundary-aware long-term tracking that prevents mask drift over extended sequences; and adaptive state transition that closes the loop between stages for failure recovery.
Furthermore, we establish a multi-modal and multi-granular benchmark from four public surgical datasets with precise instance-level masklets. 
Extensive experiments demonstrate that UniSurgSAM achieves state-of-the-art performance in real time across all prompt modalities and granularities, providing a practical foundation for computer-assisted surgery.
Code and datasets will be available at https://jinlab-imvr.github.io/UniSurgSAM.
\end{abstract}


\maketitle

\section{Introduction}
Precise surgical video segmentation is a fundamental task in computer-assisted surgery~\citep{moglia2021systematic,madani2022artificial}. 
By providing real-time localization and boundary delineation of instruments and anatomical structures, it enables critical downstream tasks such as surgical education, intraoperative decision-making support, surgical skill assessment, and the next generation of autonomous robots~\citep{jin2022exploring,ahmed2024deep,long2025surgical,pei2025instrument,pei2026synergistic}.
As surgical AI shifts toward human-AI collaboration, where AI serves as a cognitive copilot~\citep{schmidgall2025will}, flexible interaction becomes essential: surgeons need to dynamically specify targets to segment throughout extended procedures, using heterogeneous cues such as visual selections, textual expressions, or audio instructions.
However, traditional surgical video segmentation methods are typically trained on a fixed set of predefined categories without flexible interaction ability~\citep{Allan2019EndoVis17, Allan2020EndoVis18}, which limits their ability to follow user-specified targets during dynamic and complex procedures.

To address this limitation, the Promptable Video Object Segmentation (PVOS) paradigm has emerged, where user-provided prompts specify the target to segment and track~\citep{zou2023segment}. 
However, existing surgical adaptations address only one prompt modality in isolation, failing to meet the diverse demands of clinical practice: 
visual PVOS methods~\citep{surgicalsam2,yin2025memory,ma2025medsam2} enable precise spatial localization via points or boxes but lack linguistic interaction, 
while linguistic PVOS methods~\citep{wang2024video,wei2026moves,resurgsam2} support natural language specification but remain confined to textual input without audio support for hands-free interaction. 
Consequently, there is a critical need for a unified model that supports visual, textual, and audio prompts, accommodating diverse clinical scenarios while maintaining high-precision surgical video segmentation.

Beyond unifying prompt modalities, designing such a model raises a fundamental architectural challenge. Since real-time surgical assistance demands online processing that handles each frame sequentially, this naturally leads to a two-stage formulation: a target initialization stage that converts the given prompt into an initial mask, and a tracking stage that propagates the mask for temporal segmentation~\citep{sam2,cuttano2025samwise,resurgsam2}.
However, these two stages inherently exhibit \textbf{optimization interference}: the initialization stage requires understanding \textit{what} to segment from the given prompt, demanding high-level semantic interpretation, whereas the tracking stage emphasizes \textit{where} the target moves, requiring spatial coherence under motion, occlusion, and appearance changes. 
Existing coupled two-stage frameworks~\citep{sam2,cuttano2025samwise,resurgsam2} overlook this interference by entangling both objectives within a single decoder, compromising both initialization accuracy and long-term tracking stability.
Therefore, we propose a \textbf{decoupled two-stage framework} with independently optimized decoders for each stage, resolving this interference and providing a stable foundation for unified surgical PVOS.

\begin{figure}
    \centering
    \includegraphics[width=\columnwidth]{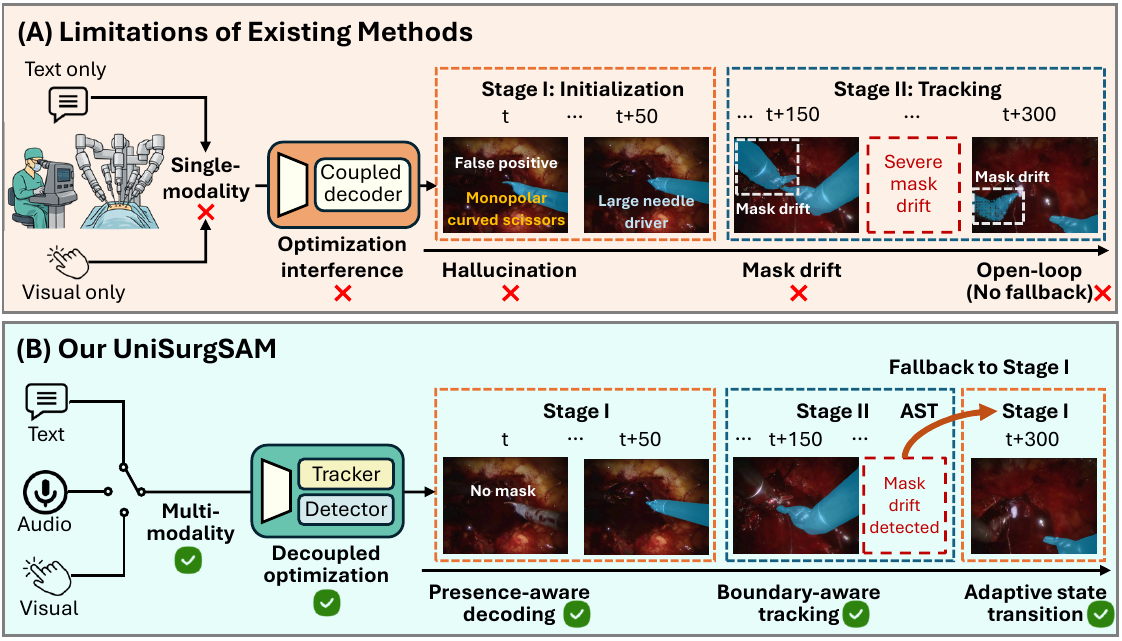}
    \caption{
    Motivation.
    (A) Existing single-modal PVOS methods rely on a coupled decoder that causes optimization interference, with fragile open-loop pipelines prone to hallucinations and mask drift.
    (B) UniSurgSAM achieves unified PVOS through the decoupled two-stage framework for stable optimization, presence-aware decoding to suppress hallucinations, boundary-aware tracking to prevent mask drift, and adaptive state transition for closed-loop failure recovery.
    \textit{Textual prompt for the example: ``large needle driver is manipulating tool on the right.'' In (A) at time $t$, the model hallucinates a mask on the wrong object when the target is absent.}
    }
    \label{fig:overview}
\end{figure}

While this decoupled framework provides a stable optimization foundation, achieving reliable surgical PVOS further requires addressing the challenges within each stage and across stages.
For Stage I \textbf{initialization}, surgical instruments frequently enter and exit the field of view. Visual prompts inherently confirm target existence through spatial interaction, but linguistic prompts lack such confirmation, making them vulnerable to hallucinations that predict false-positive masks when the target is absent. Yet existing methods rarely model target absence~\citep{li2023robust, wu2024toward}, leaving such hallucinations potentially misleading surgical decision-making.
For Stage II \textbf{tracking}, blood, smoke, and rapid view shifts obscure object boundaries, causing gradual mask drift that progressively deviates from the true target. 
Moreover, standard memory strategies fail to retain critical historical information over the extended duration of surgical procedures~\citep{yin2025memory}.
For \textbf{coordination}, linguistic PVOS is particularly fragile to cascading failures, as linguistic prompts lack the explicit spatial anchoring of visual interactions, making initialization credibility uncertain and tracking drift unrecoverable. Yet the persistent semantic descriptions provide a natural anchor for both verifying initialization credibility and correcting tracking drift, forming a closed-loop system that existing methods fail to exploit.

To address these challenges, we introduce three key designs within the decoupled two-stage framework to ensure reliability.
For \textbf{initialization}, 
Reliable Presence-Aware Decoding (RPAD) explicitly models target absence to suppress hallucinations.
For \textbf{tracking}, Boundary-Aware Long-Term Tracking (BLT) prevents mask drift through geometric constraints and diversity-driven memory. 
For \textbf{coordination}, Adaptive State Transition (AST) closes the loop between stages, validating initialization credibility before activating tracking and triggering re-initialization upon detecting drift.

Our main contributions are summarized as follows:
\begin{enumerate}
\item We propose UniSurgSAM, a unified PVOS model that supports visual, textual, and audio prompts for flexible and reliable surgical video segmentation.
\item We introduce a decoupled two-stage framework that resolves the optimization interference between target initialization and tracking through independently optimized decoders.
\item We design three complementary mechanisms to ensure reliability: presence-aware decoding for hallucination suppression, boundary-aware long-term tracking for geometric precision and temporal consistency, and adaptive state transition for closed-loop failure recovery.
\item We establish a comprehensive multi-modal and multi-granular benchmark across four surgical datasets with spatio-temporal masks, diverse prompts, and part-level annotations, on which UniSurgSAM achieves state-of-the-art performance in real time.
\end{enumerate}

This work substantially extends our preliminary version presented at MICCAI 2025~\citep{resurgsam2}, generalizing from text-only referring segmentation to unified multi-modal and multi-granular surgical PVOS. The key extensions are as follows:
(1) We upgrade the two-stage design from a coupled to a \textbf{decoupled framework}, resolving the optimization interference unaddressed in ReSurgSAM2.
(2) We enhance both stages and their coordination with \textbf{presence-aware decoding}, \textbf{boundary-aware long-term tracking}, and \textbf{consensus-based fallback} for reliable long-term segmentation.
(3) We expand the benchmark from two to four datasets, adding Uni-RARP50 for long-duration tracking and Uni-SurgAI3.8K for anatomical tissue deformation, with all datasets unified into a consistent masklet format enriched with diverse prompt modalities and fine-grained part-level annotations.
(4) Extensive experiments, ablation studies, and in-depth analyses validate each proposed component and demonstrate that UniSurgSAM achieves state-of-the-art performance across all four datasets, prompt modalities, and granularities.

\section{Related Work}
\subsection{Promptable Video Object Segmentation}
Video Object Segmentation (VOS) aims to track and segment specific objects across video sequences~\citep{oh2019video, zhou2022survey}. 
To alleviate the labor-intensive requirement of first-frame masks in traditional semi-supervised VOS, interactive segmentation paradigms have emerged as flexible alternatives~\citep{zou2023segment}. 
In this work, we collectively refer to these as Promptable Video Object Segmentation (PVOS), broadly categorized into two branches. The first branch, Interactive VOS (iVOS) or visual PVOS, utilizes visual prompts (e.g., points or boxes), with SAM2~\citep{sam2} as a representative memory-based framework. The second branch, Referring VOS (RVOS) or textual PVOS, employs linguistic expressions to specify targets in the video~\citep{wu2022language, cuttano2025samwise}.
Despite their shared objective, these branches are often studied in isolation. Their synergistic integration and the extension to audio prompts for truly hands-free interaction remain largely unexplored.

\subsection{Surgical Video Segmentation}
In the surgical domain, video segmentation has evolved from static semantic segmentation to dynamic interactive segmentation.
Early efforts focused on semantic segmentation, from binary instrument segmentation to multi-class type and part segmentation on datasets such as EndoVis17~\citep{Allan2019EndoVis17} and EndoVis18~\citep{Allan2020EndoVis18}. 
These tasks were addressed by CNN architectures such as U-Net and Mask R-CNN~\citep{gonzalez2020in}, followed by Transformer models such as Mask2Former with enhanced global context modeling~\citep{ceron2022real,ayobi2023matis,wang2025lacoste,zhao2025rethinking}. 
While the adaptation of foundation models like SurgicalSAM~\citep{Yue2024surgsam}, TPSIS~\citep{zhou2023text}, Med-SA~\citep{wu2025medical}, and MA-SAM~\citep{chen2024ma} improved semantic precision, these methods remained constrained by predefined categories or frame-level analysis.
With the emergence of PVOS, research has shifted to interactive surgical segmentation.
Recent adaptations have branched into visual PVOS ~\citep{surgicalsam2, yin2025memory, ma2025medsam2,liu2025sam2s} and linguistic PVOS~\citep{wang2024video,resurgsam2,wei2026moves}.
However, these methods focus on single-modality interactions and lack a unified model that supports diverse prompt modalities for extended surgical procedures.

\subsection{Memory Mechanisms for Long-term Tracking}
A key challenge for these PVOS methods is maintaining temporal consistency in long-term tracking. To this end, existing VOS frameworks utilize memory banks to store historical frame information~\citep{cheng2022xmem, sam2}.
For extended surgical procedures that may last for hours, efficient memory management is crucial. 
While recent adaptations such as SAM2Long~\citep{sam2long}, DAM4SAM~\citep{dam4sam}, and MA-SAM2~\citep{yin2025memory} improve memory utilization through dynamic frame selection, and our previous work ReSurgSAM2~\citep{resurgsam2} further introduced diversity-driven memory for surgical scenarios, none of these methods explicitly model geometric boundary constraints, leading to mask drift and error propagation in long-term surgical tracking.

\begin{figure*}[!t]
    \centering
    \includegraphics[width=\textwidth]{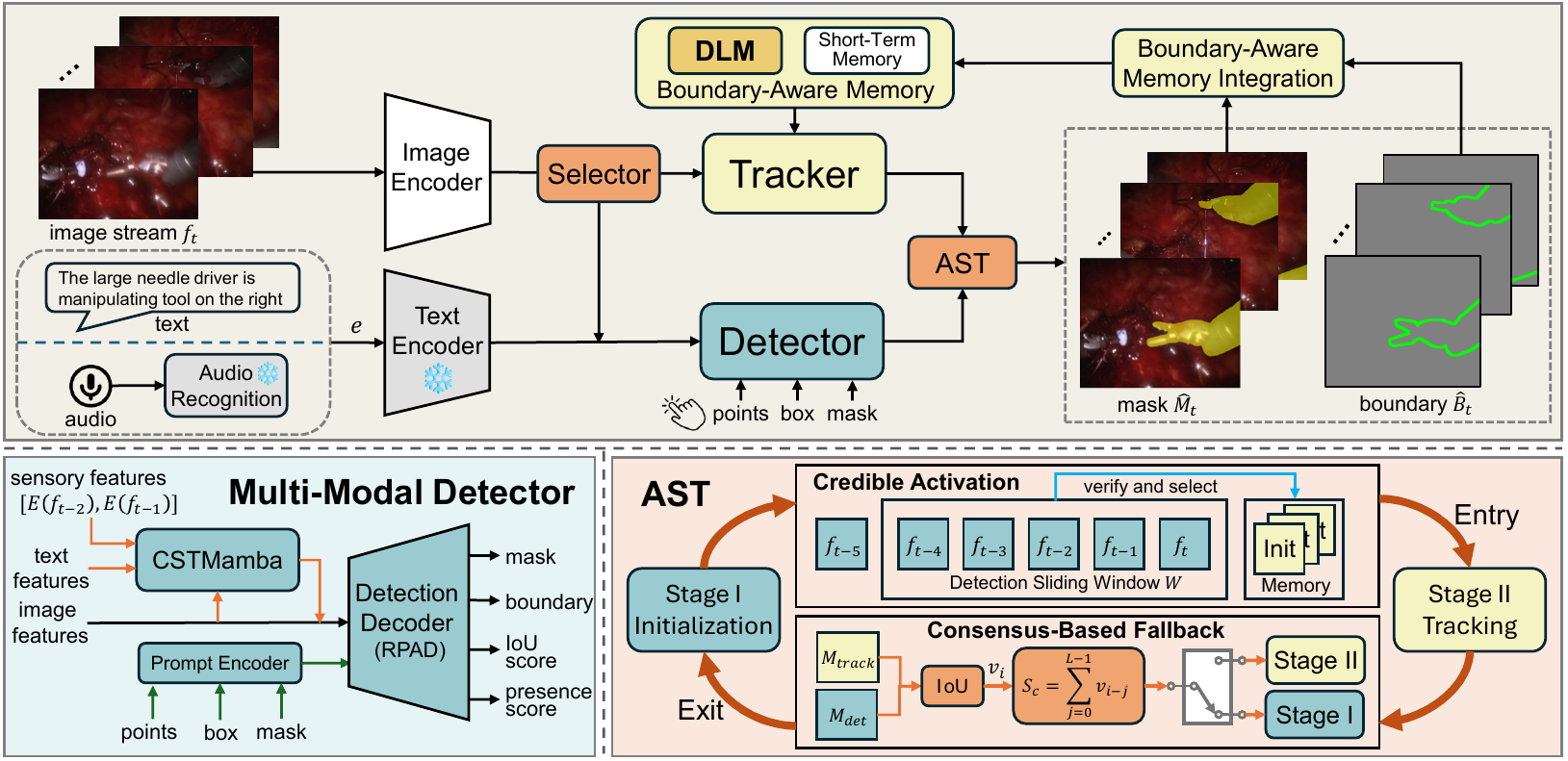}
    \caption{Overview of UniSurgSAM. The model adopts a decoupled two-stage framework that independently supports visual, textual, or audio prompts: Stage I performs promptable initialization from the given prompt, while Stage II conducts boundary-aware long-term tracking. For linguistic prompts, AST acts as a central controller that routes data to the detector or tracker via a selector, coordinating bidirectional switching through credible activation (Entry) and consensus-based fallback (Exit).}
    \label{fig:architecture}
\end{figure*}

\section{Methodology}
Given a surgical video sequence $\mathcal{V} = \{f_t\}_{t=1}^T$ and a user prompt $\mathcal{P}$, our goal is to predict a sequence of binary segmentation masks $\mathcal{\hat{M}} = \{\hat{M}_t\}_{t=1}^T$ for the intended target. 
The prompt $\mathcal{P}$ can be either a \textit{visual prompt} ($\mathcal{P}_{vis}$) such as points or box on the initial frame $f_1$, or a \textit{linguistic prompt} ($\mathcal{P}_{text}$ or $\mathcal{P}_{audio}$) in the form of a textual expression or audio instruction, where audio is first transcribed via ASR~\citep{radford2023robust}. 
Regardless of the prompt modality, the task naturally decomposes into a two-stage process: target initialization from the given prompt and tracking of the resulting mask.

\subsection{Decoupled Two-Stage Framework}
\label{sec:overview}
As shown in Fig.~\ref{fig:architecture}, UniSurgSAM adopts a decoupled two-stage framework comprising a multi-modal \textit{detector} for target initialization and a \textit{tracker} for long-term tracking. 
Both stages share an image encoder, as visual feature extraction remains largely task-agnostic~\citep{vandenhende2021multi}. 
The detector integrates cross-modal fusion with a detection decoder for semantic discrimination, while the tracker combines memory attention with a tracking decoder for tracking. 
Unlike existing coupled frameworks that entangle both objectives within a single decoder, our two decoders are independently optimized, allowing each stage to specialize without mutual interference. A selector routes the image features to the detector or tracker depending on the current system state, enabling seamless switching between initialization and tracking.

Built upon this architecture, the framework comprises two stages coordinated by a transition mechanism:
\textbf{Stage I} (Sec.~\ref{sec:init_stage}), where the \textit{detector} produces reliable initial masks from diverse prompts through presence-aware decoding;
\textbf{Stage II} (Sec.~\ref{sec:tracking}), where the \textit{tracker} propagates the initial masks through boundary-aware long-term tracking;
and \textbf{Adaptive State Transition} (Sec.~\ref{sec:transition}), which governs the selector for bidirectional switching between stages, enabling credible activation and closed-loop failure recovery.
As both decoders perform mask prediction, presence estimation, and boundary decoding, they share the same loss function:
\begin{equation}
\mathcal{L}_{total} = \lambda_{m}\mathcal{L}_{mask} + \mathcal{L}_{dice} + \mathcal{L}_{iou} + \mathcal{L}_{pre} + \lambda_{b}\mathcal{L}_{boundary},
\end{equation}
where $\mathcal{L}_{mask}$, $\mathcal{L}_{dice}$, and $\mathcal{L}_{iou}$ follow SAM2 for mask prediction and quality estimation, with $\lambda_{m}=20$ consistent with SAM2.
The boundary loss $\mathcal{L}_{boundary}$ ($\lambda_{b}=10$) enforces geometric fidelity (Sec.~\ref{sec:tracking}), and the presence loss $\mathcal{L}_{pre}$ supervises explicit presence prediction (Sec.~\ref{sec:rpad}).
Despite sharing the same supervision, each decoder emphasizes different aspects: the detection decoder primarily leverages presence modeling to suppress hallucinations during initialization, whereas the tracking decoder primarily benefits from boundary decoding to prevent mask drift during long-term tracking.

\subsection{Stage I: Unified Promptable Initialization}
\label{sec:init_stage}
Stage I accommodates both visual and linguistic prompts through the detector, with distinct processing pipelines that converge at the detection decoder for mask prediction.

For \textbf{visual prompts} ($\mathcal{P}_{vis}$), interactive inputs such as points or boxes directly specify the target on the initial frame. The prompt is encoded by the standard SAM2 prompt encoder and decoded into an initial mask. Since visual prompts provide explicit spatial constraints that inherently confirm target existence, they bypass the presence modeling and directly activate Stage II tracking.

For \textbf{linguistic prompts} ($\mathcal{P}_{text}$ or $\mathcal{P}_{audio}$), textual expressions are used directly, while audio instructions are first transcribed via ASR and normalized to correct potential transcription errors. The resulting expression $e$ is then encoded by a frozen CLIP text encoder and a trainable MLP layer to extract text features $F_{\text{text}}$.
Due to the absence of explicit spatial constraints, linguistic prompts require additional vision-language fusion and presence-aware decoding to produce reliable initial masks.

\subsubsection{Vision-Language Fusion via CSTMamba}
\label{sec:cstmamba}
For linguistic prompts, we adopt Cross-Modal Spatial-Temporal Mamba (CSTMamba)~\citep{resurgsam2} to fuse text features $F_{\text{text}}$ with current frame features $E(f_t)$ and sensory features $[E(f_{t-2}), E(f_{t-1})]$, where $E(\cdot)$ denotes the image encoder.
We stack three CSTMamba blocks, each integrating STMamba~\citep{yang2025vivim} with bidirectional cross-modal attention, where STMamba leverages the selective scan mechanism to capture long-range spatio-temporal dependencies with linear complexity, and the cross-modal attention enables vision-language interaction for semantic alignment.
This produces fused image features $E'(f_t)$ semantically aligned with the query, enabling the detection decoder to accurately localize and segment the target.

\subsubsection{Reliable Presence-Aware Decoding (RPAD)}
\label{sec:rpad}
To ensure clinical reliability, we equip the detection decoder with explicit \textit{target presence modeling} to suppress hallucinations, and \textit{multi-granularity decoding} to accommodate targets at different hierarchical levels, from complete instruments to specific components.

\paragraph{\textbf{Presence Modeling.}}
To handle target disappearance in surgical videos, we extend SAM2's occlusion prediction mechanism to the initialization stage. 
Specifically, we employ a presence head (a lightweight MLP) that predicts a presence score $s^{p} \in [0, 1]$. 
While SAM2 uses this head only during tracking to detect occlusions, we equip both decoders with this head and further leverage it in Stage I to explicitly determine whether the query target exists in the current frame, suppressing hallucinated detections before activating the tracking stage.

To further enforce robust presence discrimination, we introduce a \textit{Negative Sampling Strategy} during training by constructing:
(1) \textit{category-level negatives}, created by querying objects that are absent in the current clip, with randomized spatial attributes; 
(2) \textit{spatial hard negatives}, created by inverting directional descriptors for categories with a single instance (e.g., querying ``left" when that instrument only appears on the ``right").
For these negative samples, the presence score is supervised toward zero (indicating absence) via binary cross-entropy $\mathcal{L}_{pre}$, thereby suppressing false-positive hallucinations.

\paragraph{\textbf{Granularity-Aware Decoding.}}
Beyond presence modeling, surgical procedures often require tracking objects at different granularities, from complete instruments to specific components.
To accommodate this, we extend SAM2's multi-output design to linguistic prompts, where the decoder predicts three candidate masks at hierarchical levels (whole, part, subpart) consistent with SAM2's visual prompting interface, while enabling users to track either complete instruments (e.g., a ``grasper'') or specific components (e.g., its ``tip'').
During training, an IoU head estimates the quality score $s^{iou}$ for each candidate following a ``winner-takes-all'' strategy, optimizing only the mask with the minimum loss. 
To maintain estimator calibration, gradients for the IoU head are suppressed when the target is absent. 
During inference, the mask with the highest $s^{iou}$ is selected. Both $s^{iou}$ and $s^{p}$ are then used to determine object presence and enable tracking activation (Sec.~\ref{sec:activation}).

\subsection{Stage II: Boundary-Aware Long-Term Tracking}
\label{sec:tracking}
Upon activation via visual interaction or credible activation, the system transitions to the tracking stage. 
The tracker builds on SAM2~\citep{sam2}, retrieving relevant features from a memory bank via memory attention. 
However, in surgical scenarios, boundaries are frequently obscured by blood, smoke, and rapid view shifts, causing gradual mask drift where predicted contours progressively deviate from the true target. 
Moreover, standard memory strategies lack long-term diversity for extended surgical videos. 
To address this, we enhance the tracker with three complementary designs: boundary decoding and boundary-aware memory integration to prevent mask drift through geometric constraints, and diversity-driven long-term memory for robust tracking.

\paragraph{\textbf{Boundary Decoding.}}
The tracking decoder employs a dedicated boundary head to predict a binary boundary map $\hat{B}_t$ in parallel with the mask prediction $\hat{M}_t$, supervised by a boundary-aware loss $\mathcal{L}_{boundary}$ via binary cross-entropy.
The ground truth boundary $B_t$ is generated by applying morphological gradient operations to the mask annotation $M_t$. 
This explicit boundary supervision compels the model to prioritize fine-grained edge gradients, improving contour delineation for both rigid instruments and deformable anatomy in complex environments.
Moreover, by jointly optimizing boundary and mask predictions, the boundary head acts as an auxiliary regularizer that enforces geometric consistency in the mask output, thereby maintaining precise object delineation and improving tracking stability.

\paragraph{\textbf{Boundary-Aware Memory Integration.}}
Standard memory representations rely solely on mask and appearance features, which are insufficient to distinguish targets from visually similar distractors, especially under surgical conditions where blood, smoke, and tissue deformation obscure discriminative cues.
To address this, we explicitly integrate geometric structures into the memory bank.
For each frame, the predicted boundary map $\hat{B}_t$ and mask $\hat{M}_t$ are concatenated, downsampled, and fused with image features $E(f_t)$ to generate a boundary-aware memory representation:
\begin{equation}
    H_{t} = \text{Conv}(\text{Conv}([\hat{M}_t, \hat{B}_t]) + E(f_t)),
\end{equation}
where $[\cdot]$ denotes concatenation and $\text{Conv}(\cdot)$ represents convolutional layers.
By incorporating boundary cues, the memory representation captures both semantic and geometric information, enabling more discriminative retrieval that suppresses mask drift toward adjacent structures.

\begin{figure}
    \centering
    \includegraphics[width=\columnwidth]{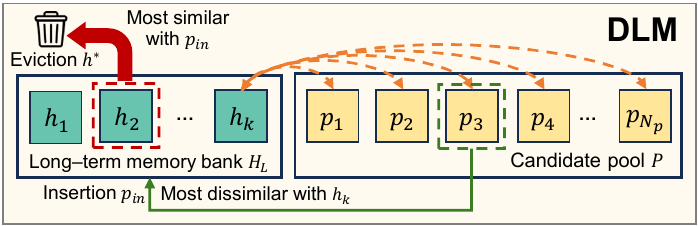}
    \caption{Illustration of Diversity-Driven Long-Term Memory.}
    \label{fig:dlm}
\end{figure}
\paragraph{\textbf{Diversity-Driven Long-Term Memory (DLM).}}
To sustain tracking over long durations, we enhance memory management beyond SAM2's FIFO strategy, which only retains recent context via a short-term memory bank, discarding critical historical information. Our preliminary work~\citep{resurgsam2} introduced an additional long-term memory bank $\mathcal{H}_L$ with bounded capacity $N_L$ and diversity-driven insertion. However, its eviction still discards the oldest frames regardless of informativeness, limiting representational breadth.
We adopt the insertion strategy from~\cite{resurgsam2} and further improve eviction.
As illustrated in Fig.~\ref{fig:dlm}, we maintain a candidate pool $P \leftarrow P \cup \{f_t \mid s^{iou}_t > \gamma_{iou}\}$ of high-confidence frames, 
where each candidate is indexed as $p_i$.
Following \cite{resurgsam2}, when $|P|$ reaches $N_p$ frames, the candidate $p_{in}$ most dissimilar to the most recently inserted frame $h_k$ in $\mathcal{H}_L$ is selected for insertion by cosine similarity in the image feature space, and $P$ is cleared. 
To improve eviction, when $\mathcal{H}_L$ reaches capacity, we replace the frame $h^*$ most similar to $p_{in}$ by cosine similarity, rather than the oldest frame, to maximize representational diversity:
\begin{equation}
    h^* = \mathop{\arg\max}_{h_j \in \mathcal{H}_L} \frac{E(p_{in}) \cdot E(h_j)}{\|E(p_{in})\| \|E(h_j)\|}.
\end{equation}
By prioritizing diversity over recency, this strategy retains informative frames across viewpoints and deformations. 
During tracking, the tracker retrieves features from the short-term memory, the long-term memory, and the initial frame memory, which together form the boundary-aware memory in Fig.~\ref{fig:architecture}, to suppress drift in extended sequences.

\subsection{Adaptive State Transition (AST)}
\label{sec:transition}
Unlike visual PVOS where targets are explicitly anchored by user clicks, linguistic PVOS lacks spatial anchoring for reliable initialization and requires mechanisms to detect tracking failures. 
AST addresses this by acting as a central controller that regulates a selector to route data between the detector and tracker through two logical gates: \textit{Entry} for credible activation and \textit{Exit} for automated fallback.

\subsubsection{Entry: Credible Activation via Temporal Verification}
\label{sec:activation}
Directly activating the tracker based on a single-frame detection is unreliable, as it may be triggered by transient false positives or momentary occlusions. 
Following ReSurgSAM2's temporal verification protocol, we buffer Stage I outputs in a sliding window of size $N_w$ and activate tracking only when all frames consistently meet high-confidence criteria:
\begin{equation}
    W=\{f_j \mid j \in [t-N_w+1, t] \land s^{iou}_j > \delta_{iou} \land s^{p}_j > \delta_{p}\},
\end{equation}
where $s^{iou}_j$ and $s^{p}_j$ denote the IoU quality and presence scores, with $\delta_{iou}$ and $\delta_{p}$ as their respective thresholds, which also serve as the detection criteria in Stage I to determine whether the queried object is present in the frame.
Upon full verification ($|W|=N_w$), unlike ReSurgSAM2's single-frame approach, we initialize Stage II using the Top-$K$ frames with the highest IoU scores from $W$, providing the tracker with richer spatio-temporal context for more robust tracking.

\subsubsection{Exit: Consensus-Based Fallback}
\label{sec:fallback}
For linguistic PVOS, once the system enters the tracking stage, tracking reliability inevitably degrades over time due to occlusions or error accumulation, causing drift toward visually similar distractors.
Since linguistic prompts persist throughout the video as semantic anchors, they can verify whether the tracked object still matches the query.

Leveraging this, we implement \textit{consensus-based fallback} that periodically validates tracking correctness via semantic constraints to form a closed-loop system.
Every $N_c$ frames, the detector is queried with the linguistic prompt to generate a semantic reference $M_{det}$. 
We compute the IoU between the tracker's prediction $M_{track}$ and $M_{det}$ for spatial consistency, and store the result $v_i$ in a queue of size $L$:
\begin{equation}
    v_i = \begin{cases} 
    +1, & \text{if IoU}(M_{track}, M_{det}) \geq \delta_c \\ 
    -1, & \text{otherwise} 
    \end{cases}
\end{equation}
where $\delta_{c}$ is the spatial consistency threshold. Once the queue is full, a fallback is triggered if the cumulative consensus score $S_c = \sum_{j=0}^{L-1} v_{i-j}$ is negative ($S_c < 0$), reverting the system to Stage I for re-initialization. 
This closed-loop design enables target re-anchoring via semantic cues, preventing the irreversible error accumulation inherent in open-loop tracking.

\section{Experiments}
\subsection{Datasets and Implementation Details}
\subsubsection{Dataset Construction}
Building upon our preliminary work~\citep{resurgsam2}, we construct a unified PVOS benchmark comprising four surgical video datasets: Uni-EndoVis17, Uni-EndoVis18, Uni-RARP50, and Uni-SurgAI3.8K (Table~\ref{tab:dataset}), all converted into the \textbf{masklet} (i.e., spatio-temporal segmentation mask) format for PVOS evaluation.
(1) Uni-EndoVis17/18: We directly adopted the refined datasets from~\cite{resurgsam2}, where the original EndoVis17~\citep{Allan2019EndoVis17} and EndoVis18~\citep{Allan2020EndoVis18} semantic labels were converted into instance-level masklets paired with textual descriptions. Notably, Uni-EndoVis18 was categorized into instrument (-I) and tissue (-T) subsets to evaluate distinct surgical targets.
(2) Uni-RARP50: Following the protocol in~\cite{resurgsam2}, we processed 16,295 frames from RARP50~\citep{psychogyios2023sar} by aggregating semantic part masks into whole-instrument masks, assigning instrument type labels according to the da Vinci system guidelines~\citep{IntuitiveSurgical2024daVinci}. These masks were then linked across frames to form instance-level masklets.
(3) Uni-SurgAI3.8K: We leveraged the inherent temporal continuity of SurgAI3.8K~\citep{zadeh2023surgai3} to reconstitute video clips from sequential frames, forming masklets for anatomical tissue tracking.

\begin{table}[!t]
  \centering
  \scriptsize
  \caption{Dataset Statistics. Avg. Dur. denotes average duration}
    \setlength{\tabcolsep}{2pt} 
    \begin{tabular}{lccccccc}
    \toprule
    Dataset & Video & Frame & \makecell{Masklet\\(whole)} & \makecell{Mask\\(whole)} & \makecell{Masklet\\(part)} & \makecell{Mask\\(part)} & \makecell{Avg.\\Dur. (s)} \\
    \midrule
      \rowcolor{gray!11}
    \multicolumn{8}{c}{\textit{Training Set}} \\
    Uni-EndoVis17 & 7 & 2100 & 20 & 4873 & 55 & 11696 & 300 \\
    Uni-EndoVis18-I & 11 & 1639 & 34 & 3787 & 78 & 7086 & 149 \\
    Uni-EndoVis18-T & 11 & 1639 & 25 & 2995 & - & - & 149 \\
    Uni-RARP50 & 44 & 13043 & 367 & 42554 & 482 & 63749 & 296 \\
    Uni-SurgAI3.8K & 36 & 2033 & 36 & 2033 & - & - & 56 \\
    \midrule
      \rowcolor{gray!11}
    \multicolumn{8}{c}{\textit{Testing Set}} \\
    Uni-EndoVis17 & 3 & 900 & 10 & 2265 & 26 & 4825 & 300 \\
    Uni-EndoVis18-I & 4 & 596 & 15 & 1384 & 33 & 2783 & 149 \\
    Uni-EndoVis18-T & 4 & 596 & 7 & 807 & - & - & 149 \\
    Uni-RARP50 & 10 & 3252 & 81 & 10656 & 115& 15501& 325 \\
    Uni-SurgAI3.8K & 15 & 1784 & 15 & 1784 & - & - & 119 \\
    \bottomrule
    \end{tabular}
  \label{tab:dataset} 
\end{table}

We further enriched these datasets with diverse prompt modalities and multi-granular annotations to support unified PVOS interaction.
First, to facilitate \textbf{diverse prompt modality} interaction, we provided a comprehensive prompt suite for all sequences:
(1) visual prompts, providing initial masks where interactive points or boxes can be simulated;
(2) textual prompts, comprising expressions for both whole-instance and part-level targets; 
and (3) audio prompts, recorded from native speakers following the textual instructions (e.g., ``segment the bipolar forceps'') to simulate hands-free control.
Second, for \textbf{multi-granular} evaluation, we extended the instrument-related datasets (Uni-EndoVis17, Uni-EndoVis18-I, and Uni-RARP50) with fine-grained part-level masklets (e.g., ``wrist of bipolar forceps''), enabling precise component tracking beyond whole-instrument segmentation.
Notably, Uni-EndoVis17 and Uni-RARP50 featured long-duration test sequences (average $\ge$300s), enabling robust assessment of long-term tracking stability.

\begin{table*}[!t]
\centering
\footnotesize
\caption{Quantitative comparison with Visual PVOS methods using 3-point prompts. \textbf{Bold} and \underline{underline} denote best and second best.}
\label{tab:visual_comparison}
\renewcommand{\arraystretch}{1.05}
\setlength{\tabcolsep}{4pt}
\begin{tabular}{lcccccccccccccccc}
\toprule
\multirow{2}{*}{Method} &\multicolumn{3}{c}{Uni-EndoVis17} &\multicolumn{3}{c}{Uni-EndoVis18-I} &\multicolumn{3}{c}{Uni-RARP50} &\multicolumn{3}{c}{Uni-EndoVis18-T} &\multicolumn{3}{c}{Uni-SurgAI3.8K} &\multirow{2}{*}{FPS} \\
\cmidrule(l{0.2em}r{0.2em}){2-4}\cmidrule(l{0.2em}r{0.2em}){5-7}\cmidrule(l{0.2em}r{0.2em}){8-10}\cmidrule(l{0.2em}r{0.2em}){11-13}\cmidrule(l{0.2em}r{0.2em}){14-16}
& $\mathcal{J}$\&$\mathcal{F}$ &$\mathcal{J}$ &$\mathcal{F}$ &$\mathcal{J}$\&$\mathcal{F}$ &$\mathcal{J}$ &$\mathcal{F}$ &$\mathcal{J}$\&$\mathcal{F}$ &$\mathcal{J}$ &$\mathcal{F}$ &$\mathcal{J}$\&$\mathcal{F}$ &$\mathcal{J}$ &$\mathcal{F}$ &$\mathcal{J}$\&$\mathcal{F}$ &$\mathcal{J}$ &$\mathcal{F}$ & \\
\midrule
SAM2~\citeyearpar{sam2} &75.9 &76.0 &75.8 &76.7 &76.5 &76.9 &80.2 &79.0 &81.3 &74.4 &79.3 &69.5 &86.9 &92.5 &81.2 &\underline{69} \\
SAM2Long~\citeyearpar{sam2long} &72.1 &72.1 &72.1 &75.4 &75.2 &75.6 &79.7 &78.5 &80.9 &74.5 &79.3 &69.6 &86.8 &92.5 &81.0 &22 \\
DAM4SAM~\citeyearpar{dam4sam} &76.2 &76.1 &76.2 &73.1 &72.8 &73.3 &77.5 &76.3 &78.6 &74.5 &79.4 &69.5 &86.8 &92.5 &81.1 &63 \\
SAMURAI~\citeyearpar{yang2024samurai} &75.3 &75.3 &75.3 &74.5 &74.2 &74.8 &77.9 &77.9 &77.8 &72.9 &77.5 &68.3 &86.6 &92.4 &80.7 &22 \\
MA-SAM2~\citeyearpar{yin2025memory} &75.5 &75.5 &75.4 &78.6 &78.3 &78.9 &82.5 &81.3 &83.7 &\underline{74.7} &\underline{79.6} &\underline{69.8} &86.7 &92.4 &81.0 &65 \\
SurgicalSAM2~\citeyearpar{surgicalsam2} &\underline{76.3} &\underline{76.3} &\underline{76.3} &\underline{79.0} &\underline{78.6} &\underline{79.3} &\underline{82.7} &\underline{81.5} &\underline{83.8} &74.5 &79.4 &69.6 &\underline{87.1} &\underline{92.6} &\underline{81.5} &\textbf{71} \\
\rowcolor{blue!11}
UniSurgSAM &\textbf{82.6} &\textbf{82.6} &\textbf{82.6} &\textbf{83.2} &\textbf{83.2} &\textbf{83.1} &\textbf{84.5} &\textbf{83.2} &\textbf{85.8} &\textbf{78.3} &\textbf{84.8} &\textbf{71.8} &\textbf{88.2} &\textbf{92.7} &\textbf{83.6} &68 \\
\bottomrule
\end{tabular}
\end{table*}

\begin{table*}[!t]
\centering
\footnotesize
\caption{Quantitative analysis of Linguistic PVOS (Textual and Audio). \textbf{Bold} and \underline{underline} denote best and second best in Textual PVOS.}
\label{tab:linguistic_comparison}
\renewcommand{\arraystretch}{1.05}
\setlength{\tabcolsep}{2.5pt}
\begin{tabular}{l>{\centering\arraybackslash}p{1.7cm}cccccccccccccccc}\toprule
\multirow{2}{*}{Method} &\multirow{2}{*}{Setting} &\multicolumn{3}{c}{Uni-EndoVis17} &\multicolumn{3}{c}{Uni-EndoVis18-I} &\multicolumn{3}{c}{Uni-RARP50} &\multicolumn{3}{c}{Uni-EndoVis18-T} &\multicolumn{3}{c}{Uni-SurgAI3.8K} &\multirow{2}{*}{FPS} \\
\cmidrule(l{0.2em}r{0.2em}){3-5}\cmidrule(l{0.2em}r{0.2em}){6-8}\cmidrule(l{0.2em}r{0.2em}){9-11}\cmidrule(l{0.2em}r{0.2em}){12-14}\cmidrule(l{0.2em}){15-17}
& &$\mathcal{J}$\&$\mathcal{F}$ &$\mathcal{J}$ &$\mathcal{F}$ &$\mathcal{J}$\&$\mathcal{F}$ &$\mathcal{J}$ &$\mathcal{F}$ &$\mathcal{J}$\&$\mathcal{F}$ &$\mathcal{J}$ &$\mathcal{F}$ &$\mathcal{J}$\&$\mathcal{F}$ &$\mathcal{J}$ &$\mathcal{F}$ &$\mathcal{J}$\&$\mathcal{F}$ &$\mathcal{J}$ &$\mathcal{F}$ & \\\midrule
ReferFormer~\citeyearpar{wu2022language} &Offline &62.5 &62.3 &62.6 &71.1 &71.0 &71.2 &81.7 &80.6 &82.8 &61.9 &69.9 &53.8 &82.5 &90.1 &74.8 &42 \\
MUTR~\citeyearpar{yan2024referred} &Offline &61.0 &60.8 &61.2 &67.6 &67.8 &67.3 &74.5 &73.3 &75.6 &63.6 &71.5 &55.6 &81.4 &89.5 &73.2 &32 \\
SurgRef~\citeyearpar{wei2026moves} &Offline &62.3 & 62.0 &62.5 &70.3 & 70.4& 70.2&76.2 & 75.3& 77.1 &65.3 & 71.5& 59.1&81.6 & 89.3& 73.9 &40 \\
RSVIS~\citeyearpar{wang2024video} &Semi-online &61.3 &61.4 &61.1 &68.4 &68.6 &68.2 &72.4 &71.7 &73.0 &65.7 &72.9 &58.5 &75.7 &83.2 &68.2 &28 \\
SAMWISE~\citeyearpar{cuttano2025samwise} &Semi-online &60.1 &59.5 &60.6 &70.6 &69.8 &71.4 &82.5 &81.7 &83.2 &64.5 &71.9 &57.0 &79.3 &88.0 &70.6 &12 \\
OnlineRefer~\citeyearpar{wu2023onlinerefer} &Online &60.3 &60.3 &60.3 &72.2 &71.9 &72.5 &73.9 &72.3 &75.4 &70.6 &77.6 &63.6 &83.0 &90.6 &75.4 &26 \\
RefSAM~\citeyearpar{li2024refsam} &Online &63.6 &63.8 &63.4 &72.9 &73.4 &72.3 &78.4 &78.3 &78.4 &71.9 &77.7 &66.1 &85.1 &91.3 &78.9 &25 \\
ResurgSAM2~\citeyearpar{resurgsam2} &Online &\underline{78.1} &\underline{78.2} &\underline{78.0} &\underline{81.1} &\underline{81.5} &\underline{80.7} &\underline{84.9} &\underline{83.8} &\underline{86.0} &\underline{75.4} &\underline{81.4} &\underline{69.4} &\underline{87.1} &\underline{92.1} &\underline{82.1} &\textbf{58} \\
\rowcolor{blue!11}
UniSurgSAM (Textual) &Online &\textbf{82.4} &\textbf{82.3} &\textbf{82.4} &\textbf{85.1} &\textbf{85.3} &\textbf{84.9} &\textbf{90.5} &\textbf{89.4} &\textbf{91.6} &\textbf{79.8} &\textbf{86.2} &\textbf{73.4} &\textbf{88.2} &\textbf{92.7} &\textbf{83.7} &\underline{55} \\
\midrule
\rowcolor{green!6}
UniSurgSAM (Audio) &Online & 79.1 & 79.0 & 79.2 &83.2 & 83.4 & 82.9 &89.8 &88.7 &90.8  &79.7 &86.0 &73.3 &88.2 &92.7 &83.7 &53 \\
\bottomrule
\end{tabular}
\end{table*}

\subsubsection{Implementation}
UniSurgSAM utilizes a pretrained SAM2~\citep{sam2} with a Hiera-base+ backbone at $512\times512$ resolution to balance efficiency and performance.
We train on each dataset individually with a two-step paradigm: joint training of the detector and tracker, followed by tracker fine-tuning. Training uses the AdamW optimizer with a learning rate of 5e-5, batch size of 28, and 60 epochs on four A6000 GPUs.

\noindent{\textbf{Step I: Joint Training.}}
The detector and tracker are jointly trained to learn initialization and tracking in an end-to-end manner. 
For visual PVOS, 1 frame is used for initialization, followed by 7-frame temporal propagation. For linguistic PVOS, 3 frames are used for initialization, followed by 7 frames for tracking. All modules are trainable except the CLIP text encoder, and negative prompts are introduced with a 1:2 positive-to-negative ratio for presence modeling.

\noindent{\textbf{Step II: Tracker Fine-tuning.}}
We freeze all components except the tracker and fine-tune it with the temporal sequence configurations as Step I to enhance tracking capability.

Inference is conducted on a single NVIDIA A6000 GPU. 
For visual prompts, we follow the SAM2 protocol \citep{sam2} with a 3-point configuration at the target's first appearance.
For linguistic prompts, unlike RSVIS~\citep{wang2024video} providing text prompts only for frames where the target is present, we follow ReSurgSAM2~\citep{resurgsam2} to adopt one-time global referring expressions that remain valid throughout the video, enabling consistent segmentation despite target disappearance or reappearance.
For system hyperparameters, we adopt the configurations from our preliminary work~\citep{resurgsam2} for credible activation and memory management.
For the consensus-based fallback, we set the check interval $N_c=5$, sliding window length $L=5$, and confidence threshold $\delta_c=0.5$ to ensure efficient drift detection.

\subsubsection{Evaluation Metrics}
Following VOS protocols~\citep{cuttano2025samwise,sam2}, we adopt region similarity ($\mathcal{J}$) (i.e., IoU), boundary accuracy ($\mathcal{F}$) (i.e., boundary F-score), and their mean ($\mathcal{J}\&\mathcal{F}$) as primary metrics. 
For \textit{\textbf{visual PVOS}}, we follow SurgicalSAM2~\citep{surgicalsam2} to compute metrics from the frame where the prompt is given. 
For \textit{\textbf{linguistic PVOS}}, evaluation is conducted from the first frame of the video regardless of object presence, following ~\cite{resurgsam2}. 
To assess presence-aware decoding, we additionally report Precision and False Positive Rate (FPR), which evaluate the model's ability to avoid hallucinations when the target is absent.
Inference speed is reported in Frames Per Second (FPS).

\subsection{Comparison with State-of-the-Arts}
We conduct comprehensive experiments to validate UniSurgSAM's effectiveness across visual, textual, and audio prompts, as well as its capability for fine-grained part segmentation.
In all comparisons, we report instrument datasets (Uni-EndoVis17, Uni-EndoVis18-I, Uni-RARP50) followed by tissue datasets (Uni-EndoVis18-T, Uni-Surg AI3.8K).
\textit{All baselines use the official implementations and are trained on the same datasets for fair comparison}.

\begin{figure*}[!t]
    \centering
    \includegraphics[width=\textwidth]{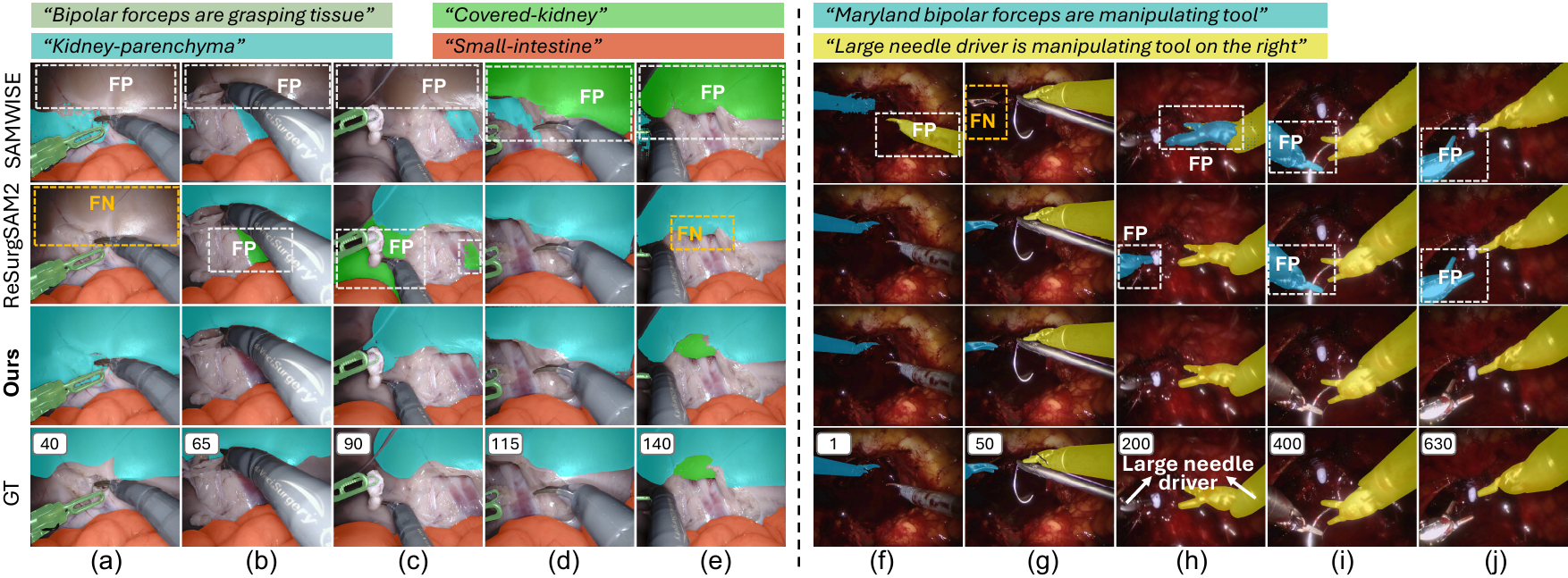}
    \caption{Qualitative comparison in Uni-EndoVis18 (left) and Uni-RARP50 (right) for Textual PVOS. FP and FN denote false positive and false negative predictions, respectively. Numbers in images denote time in seconds.}
    \label{fig:text_comparison}
\end{figure*}

\begin{figure}
    \centering
    \includegraphics[width=\columnwidth]{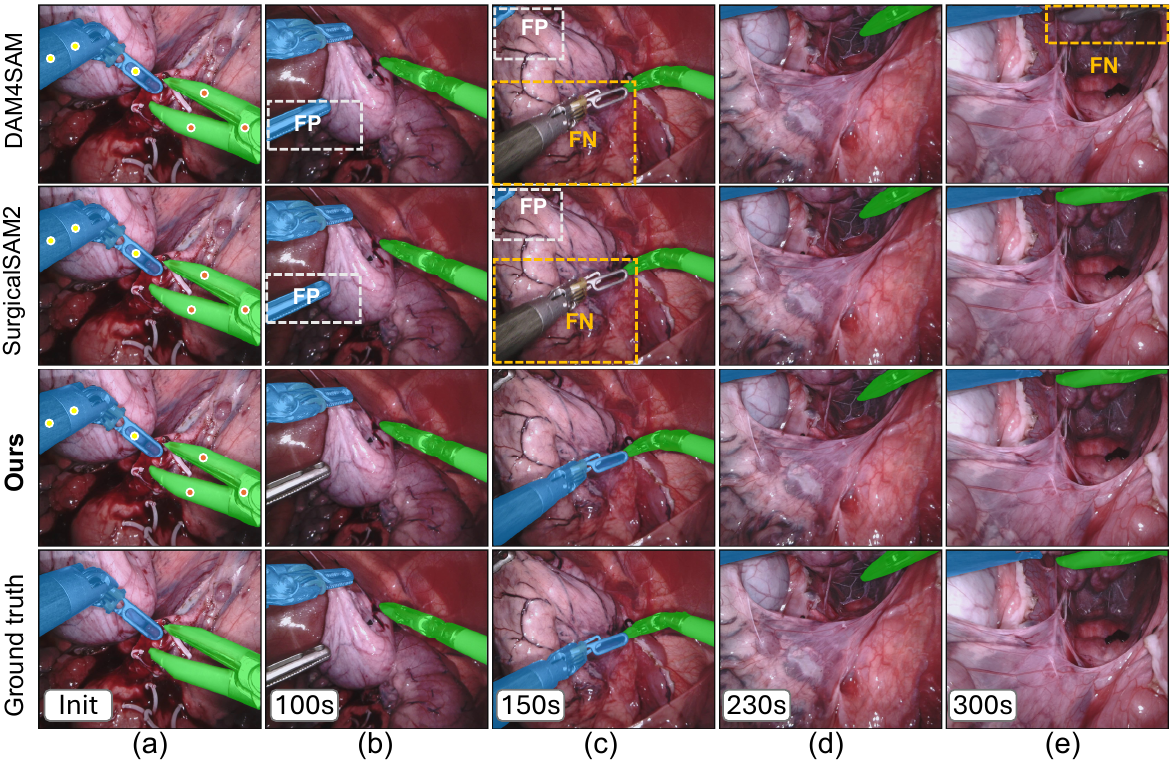}
    \caption{Qualitative comparison in Uni-EndoVis17 for Visual PVOS under three-point initialization.}
    \label{fig:visual_comparison}
\end{figure}

\subsubsection{Visual PVOS}
Table~\ref{tab:visual_comparison} presents comparisons with visual PVOS methods fine-tuned on surgical data, including SAM2 \citep{sam2}, SAM2Long \citep{sam2long}, DAM4SAM \citep{dam4sam}, SAMURAI \citep{yang2024samurai}, MA-SAM2 \citep{yin2025memory}, and SurgicalSAM2 \citep{surgicalsam2}, all using the Hiera-base+ backbone for fair comparison. Among these, MA-SAM2 and SurgicalSAM2 are surgical-specific adaptations.

Among baselines, SurgicalSAM2 and MA-SAM2 achieve the strongest overall performance through surgical-specific adaptations, while memory-augmented methods (SAM2Long, DAM4SAM, SAMURAI) show limited gains or even degradation compared to SAM2, suggesting that generic memory strategies do not transfer well to surgical scenarios.
UniSurgSAM achieves consistent improvements over the strongest baselines: +6.3 $\mathcal{J}$\&$\mathcal{F}$ on Uni-EndoVis17, +4.2 on Uni-EndoVis18-I, +1.8 on Uni-RARP50, +3.6 on Uni-EndoVis18-T, and +1.1 on Uni-SurgAI3.8K, validating the effectiveness of the decoupled framework and boundary-aware long-term tracking.
UniSurgSAM also shows significant improvement on tissue segmentation in Uni-EndoVis18-T (+3.6), where boundary-aware tracking is critical for delineating deformable anatomy.

Fig.~\ref{fig:visual_comparison} provides a qualitative comparison.
DAM4SAM introduces distractor-aware memory but lacks boundary awareness, causing frequent tracking drift between visually similar instruments, marked as FP (false positive) and FN (false negative) in Fig.~\ref{fig:visual_comparison}(b, c, e).
SurgicalSAM2 improves robustness under occlusions but still struggles during large camera movements due to limited memory capacity, as in Fig.~\ref{fig:visual_comparison}(b, c).
In contrast, UniSurgSAM maintains robust tracking by leveraging boundary-aware memory for geometric precision and diversity-driven memory across varying viewpoints, preventing both boundary drift and target loss.

\subsubsection{Textual PVOS}
\label{sec:textual_pvos}
We compare UniSurgSAM with state-of-the-art textual PVOS methods across three inference paradigms: \textbf{offline methods} (ReferFormer~\citep{wu2022language}, MUTR~\citep{yan2024referred}, SurgRef~\citep{wei2026moves}) that process all frames simultaneously, \textbf{semi-online methods} (RSVIS~\citep{wang2024video}, SAMWISE~\citep{cuttano2025samwise}) that use clip-level processing, and \textbf{online methods} (OnlineRefer~\citep{wu2023onlinerefer}, RefSAM~\citep{li2024refsam}, ReSurgSAM2~\citep{resurgsam2} with Hiera-base+ backbone) that perform frame-by-frame inference.
Among these, SurgRef, RSVIS, and ReSurgSAM2 are surgical-specific adaptations, SAMWISE extends SAM2 for textual PVOS, while the others originate from the general vision domain.

As shown in Table~\ref{tab:linguistic_comparison}, offline methods achieve stable performance by processing 64 frames concurrently but sacrifice real-time capability, and without negative sampling these models struggle to discriminate between different object categories, leading to hallucinations and performance degradation when the target is absent.
Semi-online methods face limitations: RSVIS struggles with long-term tracking due to reliance on short-term information, while SAMWISE underperforms significantly (22.3 $\mathcal{J}$\&$\mathcal{F}$ lower than ours on Uni-EndoVis17) due to inadequate surgical adaptation and suboptimal state management that causes error propagation.
Among online methods, OnlineRefer and RefSAM benefit from frame-by-frame inference but lack surgical-specific adaptations, leading to degraded performance on long-duration datasets such as Uni-EndoVis17 and Uni-RARP50.
While ReSurgSAM2 delivers the strongest baseline performance overall, it remains limited by hallucination issues and open-loop fragility.

In contrast, UniSurgSAM achieves state-of-the-art performance at 55 FPS, demonstrating exceptional \textit{\textbf{long-term tracking}} capabilities: +4.3 $\mathcal{J}$\&$\mathcal{F}$ on Uni-EndoVis17 (avg. 300s), +5.6 on Uni-RARP50 (avg. 325s), and +4.4 for tissue and +4.0 for instrument on Uni-EndoVis18, compared to ReSurgSAM2.
These improvements validate our key designs: presence-aware decoding for hallucination suppression, boundary-aware memory for consistent tracking, and adaptive state transition for resilient recovery.

Fig.~\ref{fig:text_comparison} shows qualitative comparisons.
Both SAMWISE and ReSurgSAM2 exhibit boundary drift, particularly on deformable tissues where masks gradually leak into adjacent regions, as shown in Fig.~\ref{fig:text_comparison}(a--e).
Both methods also lack robust presence modeling, causing hallucinations that misidentify absent instruments as present, as shown by the FP predictions in Uni-RARP50 (Fig.~\ref{fig:text_comparison}(f--j)).
ReSurgSAM2 further suffers from irreversible tracking drift once errors occur due to the lack of fallback mechanisms, as shown in Fig.~\ref{fig:text_comparison}(h--j).
In comparison, UniSurgSAM maintains precise contours through boundary-aware long-term tracking, suppresses hallucinations through presence-aware decoding, and recovers from drift via consensus-based fallback.

\subsubsection{Audio PVOS}
To validate UniSurgSAM's versatility across input modalities, we evaluate audio PVOS performance in Table~\ref{tab:linguistic_comparison}. 
Audio prompts were recorded by native speakers following the textual instructions (e.g., ``Segment the large needle driver on the left'') and processed through the same pipeline as textual prompts after ASR transcription. 
As no prior methods support audio prompting, we report UniSurgSAM's results for reference.
Results show comparable performance between audio and textual PVOS across all datasets, with minor degradation on certain datasets (e.g., 3.3 $\mathcal{J}$\&$\mathcal{F}$ lower on Uni-EndoVis17). This gap arises because audio prompts are transcribed via ASR into simplified expressions that lack the action descriptors present in the curated textual prompts, introducing slight discrepancies at the model input. Despite this, the overall performance demonstrates robustness to prompt modality variations and potential for hands-free surgical interaction where direct text input may be impractical.

\begin{table}[!t]
    \centering
    \footnotesize
    \caption{
    Instrument part segmentation under Textual PVOS.
    }
    \label{tab:part_segmentation} 
    \setlength{\tabcolsep}{2pt}
    \begin{tabular}{l ccc ccc ccc}
    \toprule
    \multirow{2}*{Method} & \multicolumn{3}{c}{Uni-EndoVis17} & \multicolumn{3}{c}{Uni-EndoVis18-I} & \multicolumn{3}{c}{Uni-RARP50}\\
    \cmidrule(l{0.2em}r{0.2em}){2-4}\cmidrule(l{0.2em}r{0.2em}){5-7}\cmidrule(l{0.2em}){8-10}
    ~ & $\mathcal{J}$\&$\mathcal{F}$ & $\mathcal{J}$ & $\mathcal{F}$ & $\mathcal{J}$\&$\mathcal{F}$ & $\mathcal{J}$ & $\mathcal{F}$ & $\mathcal{J}$\&$\mathcal{F}$ & $\mathcal{J}$ & $\mathcal{F}$ \\ 
    \midrule
    ReferFormer & 60.2 & 57.5 & 62.8 & 64.3 & 63.0 & 65.6 & 72.7 & 72.0 & 73.3\\
    SAMWISE & 63.7 & 60.1 & 67.2 & 67.6 & 65.4 & 63.2 & 77.9 & 77.5 & 78.3\\
    RefSAM & 66.0 & 64.6 & 67.4 & 69.3 & 67.8 & 66.2 & 74.7 & 74.2 & 75.1\\
    ReSurgSAM2 & 69.1 & 67.1 & 71.0 & 71.7 & 69.5 & 73.8 & 78.7 & 77.4 & 79.9\\
    \rowcolor{blue!11}
    UniSurgSAM & \textbf{77.7} & \textbf{75.1} & \textbf{80.3} & \textbf{77.3} & \textbf{75.3} & \textbf{79.3} & \textbf{88.4} & \textbf{87.6} & \textbf{89.1}\\
    \bottomrule
    \end{tabular}%
\end{table}

\subsubsection{Part-Level Segmentation}
Beyond whole-object segmentation, we evaluate fine-grained part segmentation under textual PVOS, where part-level queries specify both the instrument and its component (e.g., ``wrist of bipolar forceps''). Representative baselines are trained on datasets with multi-granular annotations covering whole instruments and constituent parts. 
As shown in Table~\ref{tab:part_segmentation}, UniSurgSAM achieves substantial improvements over all baselines across all datasets: +8.6 $\mathcal{J}$\&$\mathcal{F}$ over ReSurgSAM2 on Uni-EndoVis17, +5.6 on Uni-EndoVis18-I, and +9.7 on Uni-RARP50.
These gains confirm that UniSurgSAM seamlessly handles multi-granular segmentation without architectural modifications, benefiting from reliable presence-aware decoding that suppresses hallucinations under target absence and hierarchical granularity selection.

\subsubsection{Inference Efficiency}
UniSurgSAM achieves real-time inference across all modalities on a single NVIDIA A6000 GPU: 68 FPS for visual PVOS, 55 FPS for textual PVOS, and 53 FPS for audio PVOS, all exceeding real-time requirements. 
The difference in speed stems from the additional computational overhead of CSTMamba fusion and ASR transcription in the linguistic pipeline.
These speeds outperform most comparison methods while delivering superior accuracy, demonstrating a balance between efficiency and performance for practical surgical applications.

\subsection{Ablation Study}
We conduct comprehensive ablation studies to validate the effectiveness of each proposed component. 
All ablations are performed on Uni-EndoVis17, Uni-EndoVis18-I, and Uni-EndoVis18-T under the textual PVOS setting.

\subsubsection{Main Components}

Table~\ref{tab:ablation_main} presents ablation results for the main components of UniSurgSAM: decoupled decoders, Reliable Presence-Aware Decoding (RPAD), Boundary-Aware Long-Term Tracking (BLT), and Adaptive State Transition (AST).
The baseline adopts a coupled two-stage framework with CSTMamba for cross-modal feature fusion and a shared decoder for both initialization and tracking, where tracking is activated upon the first high-confidence detection.
Decoupled decoders bring consistent improvements (+2.5 $\mathcal{J}$\&$\mathcal{F}$ on Uni-EndoVis17) by separating optimization on initialization from tracking.
Incorporating RPAD with negative sampling yields further gains (+3.2 on Uni-EndoVis17, +1.9 on Uni-EndoVis18-I) by learning to reject absent targets and reduce false positives.
To isolate the contributions of BLT and AST, we evaluate them separately on top of the RPAD baseline.
BLT improves tracking consistency through boundary-aware memory, with notable gains (+2.1 on Uni-EndoVis18-T) on deformable tissues where boundary supervision is critical. 
However, without AST, the tracker may be activated by unreliable initializations, limiting overall gains.
Conversely, AST ensures credible tracking activation and enables automatic recovery from tracking failures (+4.0 on Uni-EndoVis17), but without boundary-aware memory, tracking precision degrades over long sequences.
Their combination achieves the best performance (+7.9 over the RPAD baseline on Uni-EndoVis17), effectively synergizing geometric preservation with resilient failure recovery for reliable long-term tracking.
\begin{table}[!t]
    \centering
    \footnotesize
    \caption{Ablation study of the main components ($\mathcal{J}$\&$\mathcal{F}$).}
    \label{tab:ablation_main}
    \setlength{\tabcolsep}{1.5pt}
    \begin{tabular}{ccccccc}
    \toprule
    \makecell{Decoupled\\Decoders} & \makecell{RPAD} & \makecell{BLT} & \makecell{AST} & \makecell{Uni-\\EndoVis17} & \makecell{Uni-\\EndoVis18-I} & \makecell{Uni-\\EndoVis18-T} \\
    \midrule
    & & & & 68.8 & 74.5 & 72.2 \\ 
    $\surd$ & & & & 71.3 & 76.8 & 73.5 \\
    $\surd$ & $\surd$ & & & 74.5 & 78.7 & 74.6 \\
    $\surd$ & $\surd$ & $\surd$ & & 77.4 & 81.5 & 76.7 \\
    $\surd$ & $\surd$ & & $\surd$ & 78.5 & 82.1 & 75.5 \\
    \rowcolor{blue!11}
    $\surd$ & $\surd$ & $\surd$ & $\surd$ & \textbf{82.4} & \textbf{85.1} & \textbf{79.8} \\
    \bottomrule
    \end{tabular}
\end{table}

\subsubsection{Boundary-Aware Tracking Components}
\begin{table}[!t]
\centering
\footnotesize
\caption{Ablation on boundary-aware tracking components ($\mathcal{J}$\&$\mathcal{F}$). \\Baseline: decoupled decoders + RPAD + AST.}
\label{tab:ablation_boundary}
\setlength{\tabcolsep}{2pt}
\begin{tabular}{lccc}
\toprule
Model Configuration & \makecell{Uni-\\EndoVis17} & \makecell{Uni-\\EndoVis18-I} & \makecell{Uni-\\EndoVis18-T} \\
\midrule
Baseline & 78.5 & 82.1 & 75.5 \\
\quad + Boundary Decoding & 79.7 & 83.5 & 78.2 \\
\quad + Memory Integration & 80.9 & 84.0 & 79.2 \\
\quad + DLM (FIFO eviction) & 81.6 & 84.5 & 79.4 \\
\rowcolor{blue!11}
\quad + DLM (full) & \textbf{82.4} & \textbf{85.1} & \textbf{79.8} \\
\bottomrule
\end{tabular}
\end{table}
To validate the architectural designs for boundary-aware long-term tracking (Stage II), we conduct a progressive ablation study starting from the baseline equipped with decoupled decoders, RPAD, and AST. 
Table~\ref{tab:ablation_boundary} details the improvements yielded by three key configurations:
(1) Boundary Decoding:
Adding the parallel boundary head brings significant improvements, especially on the tissue dataset (+2.7 $\mathcal{J}$\&$\mathcal{F}$ on Uni-EndoVis18-T). 
This confirms that explicit boundary supervision is critical for delineating deformable tissues where boundaries are often obscured by blood or deformation.
(2) Memory Integration: Encoding these boundary features into the memory bank yields additional gains on instrument datasets (+1.2 on Uni-EndoVis17) and tissues (+1.0 on Uni-EndoVis18-T). This geometric integration enables the tracker to distinguish targets from visually similar structures during retrieval, effectively preventing drift for both instruments and tissues.
(3) DLM: Adopting the long-term memory bank with diversity-driven insertion from our preliminary work~\citep{resurgsam2} already yields gains with FIFO eviction (+0.7 on Uni-EndoVis17), confirming the value of retaining diverse historical context. Further replacing FIFO with the proposed diversity-driven eviction brings additional improvements (+0.8 on Uni-EndoVis17, +0.6 on Uni-EndoVis18-I), demonstrating that diversity-driven eviction complements the insertion for more effective memory management in long-duration sequences.

\begin{table}[!t]
\centering
\footnotesize
\caption{Ablation on AST components under Textual PVOS ($\mathcal{J}$\&$\mathcal{F}$).}
\label{tab:ablation_ast}
\setlength{\tabcolsep}{1.8pt}
\begin{tabular}{l ccc}
\toprule
Model Configuration 
  & \makecell{Uni-\\EndoVis17} 
  & \makecell{Uni-\\EndoVis18-I} 
  & \makecell{Uni-\\EndoVis18-T} \\
\midrule
Baseline 
  & 77.4 & 81.5 & 76.7 \\
\quad + Credible Activation 
  & 79.8 & 83.2 & 78.1 \\ 
\rowcolor{blue!11}
\quad + Consensus Fallback 
  & \textbf{82.4} & \textbf{85.1} & \textbf{79.8} \\
\bottomrule
\end{tabular}
\end{table}

\begin{figure}
    \centering
    \includegraphics[width=\columnwidth]{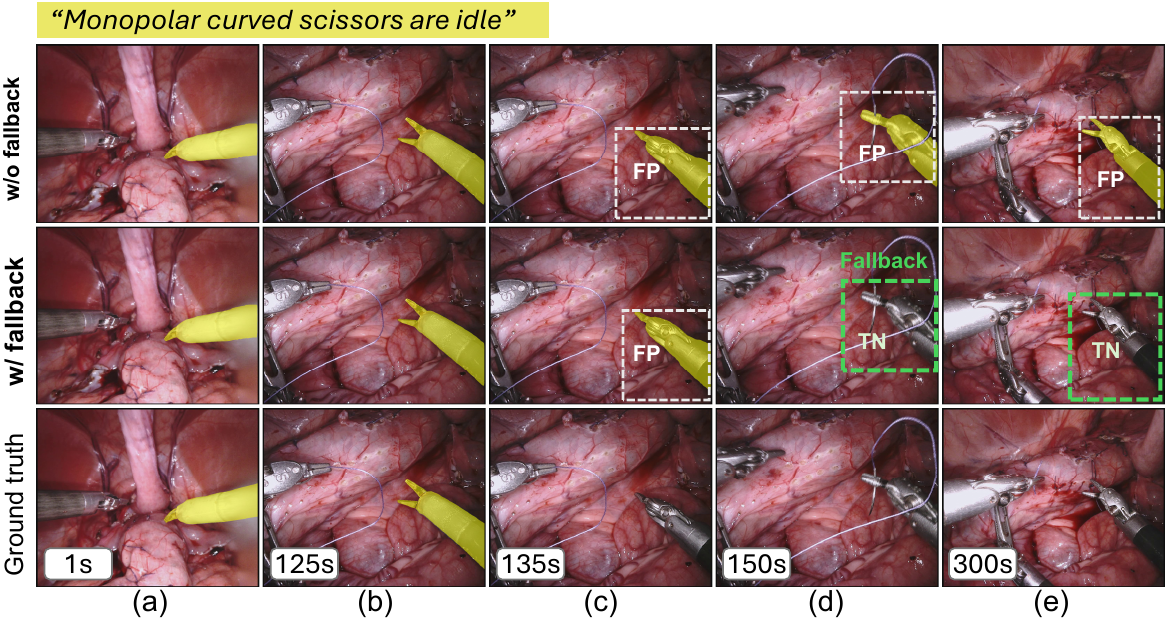}
    \caption{Qualitative analysis of consensus-based fallback in AST. The system successfully detects tracking drift and triggers re-initialization.}
    \label{fig:ast}
\end{figure}
\subsubsection{Adaptive State Transition Components}
To evaluate the contribution of the AST mechanism, we conduct a quantitative ablation on its two logic gates, as detailed in Table~\ref{tab:ablation_ast}. 
Starting from a baseline equipped with decoupled decoders, RPAD, and BLT, integrating the Entry gate for credible activation yields consistent improvements (+2.4 $\mathcal{J}$\&$\mathcal{F}$ on Uni-EndoVis17 and +1.7 on Uni-EndoVis18-I) by rejecting transient false positives to ensure reliable tracking initialization. 
Further incorporating the Exit gate for consensus-based fallback provides substantial additional gains (+2.6 on Uni-EndoVis17), which is particularly pronounced on long-duration sequences where tracking drift is more likely to accumulate over time. 
These quantitative results underscore the necessity of a closed-loop system for long-term sequences.

Fig.~\ref{fig:ast} illustrates the effect of consensus-based fallback on a representative case in Uni-EndoVis17. The detector first localizes the target and activates the tracker through credible activation (Fig.~\ref{fig:ast}(a,b)). The tracker successfully propagates the mask in subsequent frames, but drifts toward a visually similar distractor as the procedure progresses (Fig.~\ref{fig:ast}(c)). Without a fallback mechanism, this error propagates irreversibly until the end of the video. In contrast, with the Exit gate enabled, the system periodically validates tracking correctness via semantic consensus. Upon detecting inconsistency between the tracker's prediction and the detector's semantic reference, a fallback to Stage I is triggered, re-initializing detection and quickly recovering from the drift.
Together with credible activation, this fallback mechanism forms a closed-loop system that prevents irreversible error accumulation, particularly critical for extended surgical procedures where drift compounds over time.

\subsubsection{Backbone Architecture}
\begin{table}[!t]
\centering
\footnotesize
\caption{Comparison across backbone scales under Textual PVOS.}
\label{tab:backbone}
\setlength{\tabcolsep}{1.2pt}
\begin{tabular}{l ccc ccc ccc c}
\toprule
\multirow{2}{*}{Backbone} & \multicolumn{3}{c}{Uni-EndoVis17} & \multicolumn{3}{c}{Uni-EndoVis18-I} & \multicolumn{3}{c}{Uni-EndoVis18-T} & \multirow{2}{*}{FPS} \\
\cmidrule(lr){2-4}\cmidrule(lr){5-7}\cmidrule(lr){8-10}
& $\mathcal{J}$\&$\mathcal{F}$ & $\mathcal{J}$ & $\mathcal{F}$ & $\mathcal{J}$\&$\mathcal{F}$ & $\mathcal{J}$ & $\mathcal{F}$ & $\mathcal{J}$\&$\mathcal{F}$ & $\mathcal{J}$ & $\mathcal{F}$ & \\
\midrule
Small & 80.6& 80.6&80.5 &82.5 &82.8 &82.1 &77.6 &82.9 & 72.3 & \textbf{58} \\
\rowcolor{blue!11}
Base+ & \textbf{82.4} & \textbf{82.3} & \textbf{82.4} & \textbf{85.1} & \textbf{85.3} & \textbf{84.9} & \textbf{79.8} & \textbf{86.2} & \textbf{73.4} & 55 \\
\bottomrule
\end{tabular}
\end{table}

As shown in Table~\ref{tab:backbone}, UniSurgSAM maintains strong performance across backbone scales while achieving real-time inference speeds. Scaling from Hiera-small to Hiera-base+ yields consistent improvements across all datasets (+1.8 $\mathcal{J}$\&$\mathcal{F}$ on Uni-EndoVis17, +2.6 on Uni-EndoVis18-I, +2.2 on Uni-EndoVis18-T), indicating that the framework effectively benefits from increased model capacity. The performance gain comes at a negligible efficiency cost (58$\rightarrow$55 FPS), and we therefore adopt Hiera-base+ as the default backbone to balance performance and efficiency.

\subsection{Component Analysis}
\subsubsection{Analysis of Optimization Interference}
To empirically validate the optimization interference in the coupled decoder, we analyze gradient dynamics and loss convergence behavior on Uni-EndoVis17 under the textual PVOS setting with joint training.

\begin{figure}
    \centering
    \includegraphics[width=\columnwidth]{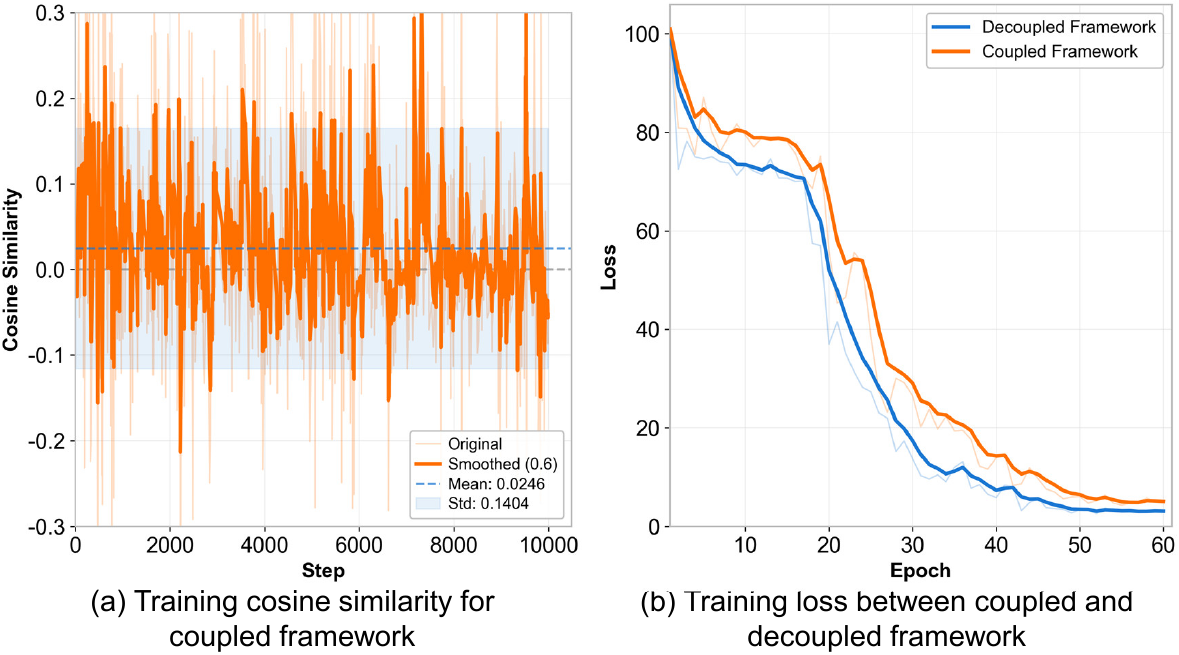}
    \caption{Analysis of optimization interference.}
    \label{fig:conflict}
\end{figure}

\paragraph{\textbf{Gradient Dynamics.}}
As shown in Fig.~\ref{fig:conflict}(a), we analyze the gradient dynamics of a coupled decoder, where the first few frames are used for initialization and the subsequent frames for tracking.
At each iteration, we sequentially compute the gradients of the losses on the target initialization frame and the subsequent tracking frames with respect to the coupled decoder parameters, and then measure their cosine similarity. 
The cosine similarity fluctuates considerably throughout training, with a near-zero mean (0.0246) and frequent sign changes, indicating that the two objectives intermittently impose conflicting optimization directions on the shared decoder parameters.

\paragraph{\textbf{Loss Convergence.}}
Fig.~\ref{fig:conflict}(b) compares the total loss curves under coupled and decoupled frameworks, both starting from the same model parameters where the detection decoder and tracking decoder share the same pretrained weights. With a coupled decoder, the total loss exhibits instability and slower convergence throughout training. In contrast, decoupled decoders allow each stage to optimize independently, yielding smoother and faster convergence. These results further explain the quantitative gains reported in Table~\ref{tab:ablation_main}, where decoupled decoders consistently improve performance across instrument and tissue segmentation.

\subsubsection{Analysis of Presence-Aware Decoding}
\begin{figure}
    \centering
    \includegraphics[width=\columnwidth]{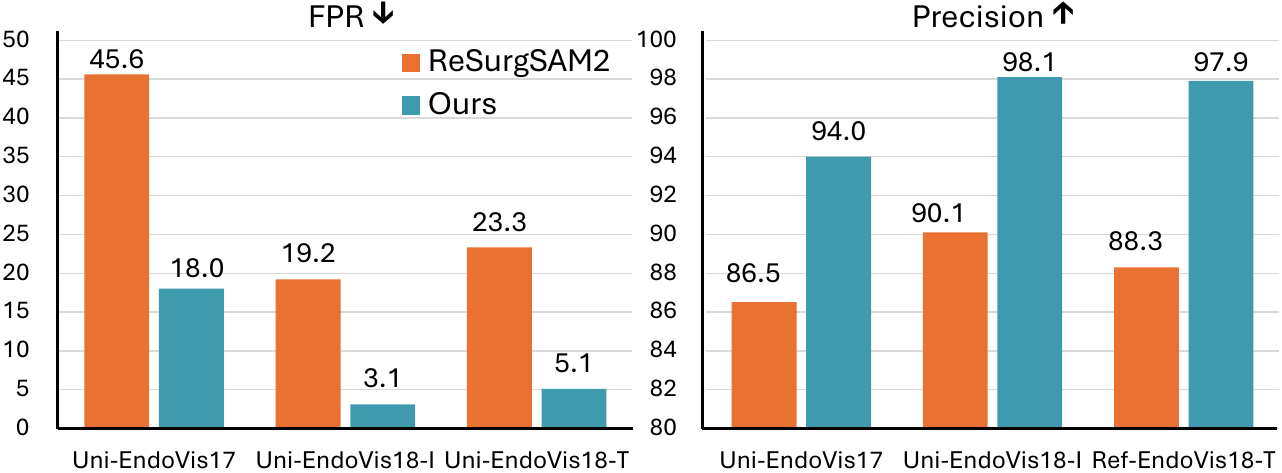}
    \caption{Analysis of presence-aware decoding with negative sampling.}
    \label{fig:presence}
\end{figure}

Beyond the $\mathcal{J}$\&$\mathcal{F}$ improvements reported in Sec.~\ref{sec:textual_pvos}, we further analyze the effectiveness of presence-aware decoding, focusing on hallucination suppression on sequences where the referred object intermittently appears and disappears.
Fig.~\ref{fig:presence} presents a detailed comparison between ReSurgSAM2 and UniSurgSAM on Uni-EndoVis17 and Uni-EndoVis18.
UniSurgSAM achieves superior hallucination suppression with substantially lower FPR and higher precision. On Uni-EndoVis17, FPR reduces from 45.6 to 18.0 while precision improves from 86.5 to 94.0.
Similar improvements are observed on Uni-EndoVis18, where FPR drops from 19.2 to 3.1 for instruments and from 23.3 to 5.1 for tissues, with precision reaching 98.1 and 97.9, respectively. 
These results validate that presence-aware decoding with negative sampling enables UniSurgSAM to accurately detect target absence, a capability essential for reliable surgical video segmentation.

\section{Discussion}
UniSurgSAM achieves state-of-the-art performance with real-time inference across all datasets, with notable gains on long-duration tracking for both instruments and tissues.
We attribute these gains to the synergy of the decoupled framework and the three reliability designs. 
Each decoder is dedicated to a single objective: the detector focuses on semantic discrimination with reliable presence modeling, while the tracker maintains geometric and temporal consistency over extended sequences through boundary-aware and diversity-driven memory.
Additionally, adaptive state transition closes the loop between stages, enabling resilient failure recovery for linguistic interaction.
Compared to our preliminary work, ReSurgSAM2~\citep{resurgsam2}, our framework yields substantial improvements on long-duration datasets while extending support to visual and audio prompts.
Clinically, the real-time capability (53+ FPS across all prompt modalities) and flexible multi-modal interaction offer practical value for intraoperative assistance.

General foundation models such as SAM3~\citep{sam3} offer broad segmentation capabilities, yet they address a fundamentally different task from surgical PVOS in terms of both formulation and efficiency.
SAM3 is designed for Promptable Concept Segmentation (PCS), detecting and segmenting \textit{all instances} of a concept. However, surgical scenarios primarily involve \textit{single-instance tracking} where surgeons monitor one specific instrument or tissue region. 
The multi-instance capability introduces unnecessary overhead, as most surgical targets appear as one or two instances that can be easily specified through simple expressions.
Computational constraints further limit clinical applicability. Our benchmark shows that SAM3 achieves \textbf{only 9 FPS} on an A6000 GPU, falling short of real-time requirements, whereas UniSurgSAM maintains \textbf{53+ FPS}. 
A direct quantitative comparison is therefore infeasible due to the fundamental difference in task formulation and the substantial domain-specific fine-tuning that SAM3 would require to operate under the surgical PVOS setting.
These limitations underscore the necessity of domain-specific frameworks tailored to the task requirements and real-time demands of surgical applications.

Despite the promising results, several limitations warrant discussion.
First, vision-language generalization is constrained to the trained surgical procedures. Training on more diverse procedure types would be needed to broaden applicability. 
Second, the ASR module is not optimized for noisy operating room environments. Incorporating domain-adapted speech recognition with noise reduction would improve audio prompt robustness.
Beyond these improvements, we will explore integration with robotic control systems to close the loop from perception to action in surgical automation.

\section{Conclusion}
This paper presents UniSurgSAM, a unified promptable model for reliable surgical video segmentation that supports visual, textual, and audio interactions.
Its decoupled two-stage framework resolves the optimization interference between initialization and tracking, enabling each stage to specialize independently.
Within this framework, presence-aware decoding suppresses hallucinations, boundary-aware long-term tracking prevents mask drift, and adaptive state transition closes the loop between stages for resilient failure recovery.
Extensive validation demonstrates state-of-the-art performance in real time across all prompt modalities and granularities, establishing a practical foundation for computer-assisted surgery.



\bibliographystyle{cas-model2-names}

\bibliography{references}

@string{PAMI = {IEEE Trans. Pattern Anal. Mach. Intell.}}

@string{CSVT = {IEEE Trans. Circuit Syst. Video Technol.}}

@string{CVPR = {Proc. IEEE/CVF Conf. Comput. Vis. Pattern Recog.}}

@string{ICCV = {Proc. IEEE/CVF Int. Conf. Comput. Vis.}}

@string{NeurIPS = {Proc. Conf. Neural Inform. Process. Syst.}}

@string{ICLR = {Proc. Int. Conf. Learn. Represent.}}

@string{AAAI = {Proc. AAAI Conf. Artif. Intell.}}

@inproceedings{wu2022language,
	title        = {Language as queries for referring video object segmentation},
	author       = {Wu, Jiannan and Jiang, Yi and Sun, Peize and Yuan, Zehuan and Luo, Ping},
	year         = 2022,
	booktitle    = CVPR,
	pages        = {4974--4984}
}

@inproceedings{yan2024referred,
	title        = {Referred by multi-modality: A unified temporal transformer for video object segmentation},
	author={Yan, Shilin and Zhang, Renrui and Guo, Ziyu and Chen, Wenchao and Zhang, Wei and Li, Hongyang and Qiao, Yu and Dong, Hao and He, Zhongjiang and Gao, Peng},
	year         = 2024,
	booktitle    = {Proc. AAAI Conf. Artif. Intell.},
	volume       = 38,
	pages        = {6449--6457}
}

@article{wang2024video,
	title        = {Video-Instrument Synergistic Network for Referring Video Instrument Segmentation in Robotic Surgery},
	author={Wang, Hongqiu and Yang, Guang and Zhang, Shichen and Qin, Jing and Guo, Yike and Xu, Bo and Jin, Yueming and Zhu, Lei},
	year         = 2024,
	journal      = {IEEE Trans. Med. Imag.},
	volume       = 43,
	number       = 12,
	pages        = {4457--4469}
}

@inproceedings{wu2023onlinerefer,
	title        = {Onlinerefer: A simple online baseline for referring video object segmentation},
	author={Wu, Dongming and Wang, Tiancai and Zhang, Yuang and Zhang, Xiangyu and Shen, Jianbing},
	year         = 2023,
	booktitle    = ICCV,
	pages        = {2761--2770}
}

@article{li2024refsam,
  title={Refsam: Efficiently adapting segmenting anything model for referring video object segmentation},
  author={Li, Yonglin and Zhang, Jing and Teng, Xiao and Zhang, Haoyu and Liu, Xinwang and Lan, Long},
  journal={Neural Netw.},
  pages={108000},
  year={2025},
  publisher={Elsevier}
}

@article{moglia2021systematic,
	title        = {A systematic review on artificial intelligence in robot-assisted surgery},
	author       = {Moglia, Andrea and Georgiou, Konstantinos and Georgiou, Evangelos and Satava, Richard M and Cuschieri, Alfred},
	year         = 2021,
	journal      = {Int. J. Surg.},
	publisher    = {Elsevier},
	volume       = 95,
	pages        = 106151
}

@article{zhou2023text,
	title        = {Text promptable surgical instrument segmentation with vision-language models},
	author       = {Zhou, Zijian and Alabi, Oluwatosin and Wei, Meng and Vercauteren, Tom and Shi, Miaojing},
	year         = 2023,
	journal      = NeurIPS,
	volume       = 36,
	pages        = {28611--28623}
}

@inproceedings{surgicalsam2,
  title={Surgical {SAM} 2: Real-time Segment Anything in Surgical Video by Efficient Frame Pruning},
  author={Haofeng Liu and Erli Zhang and Junde Wu and Mingxuan Hong and Yueming Jin},
  booktitle={Neural Information Processing Systems Workshop on Advancements In Medical Foundation Models},
  year={2024}
}

@inproceedings{cuttano2025samwise,
  title={Samwise: Infusing wisdom in sam2 for text-driven video segmentation},
  author={Cuttano, Claudia and Trivigno, Gabriele and Rosi, Gabriele and Masone, Carlo and Averta, Giuseppe},
  booktitle=CVPR,
  pages={3395--3405},
  year={2025}
}

@inproceedings{sam2long,
  title={Sam2long: Enhancing sam 2 for long video segmentation with a training-free memory tree},
  author={Ding, Shuangrui and Qian, Rui and Dong, Xiaoyi and Zhang, Pan and Zang, Yuhang and Cao, Yuhang and Guo, Yuwei and Lin, Dahua and Wang, Jiaqi},
  booktitle=ICCV,
  pages={13614--13624},
  year={2025}
}

@inproceedings{gonzalez2020in,
	title        = {Isinet: an instance-based approach for surgical instrument segmentation},
	author       = {Gonz{\'a}lez, Cristina and Bravo-S{\'a}nchez, Laura and Arbelaez, Pablo},
	year         = 2020,
	booktitle    = {Proc. Int. Conf. Med. Image Comput. Comput.-Assist. Intervent.},
	pages        = {595--605}
}

@inproceedings{Yue2024surgsam,
	title        = {Surgicalsam: Efficient class promptable surgical instrument segmentation},
	author       = {Yue, Wenxi and Zhang, Jing and Hu, Kun and Xia, Yong and Luo, Jiebo and Wang, Zhiyong},
	year         = 2024,
	booktitle    = {Proc. AAAI Conf. Artif. Intell.},
	volume       = 38,
	pages        = {6890--6898}
}

@article{Allan2019EndoVis17,
	title        = {2017 robotic instrument segmentation challenge},
	author={Allan, Max and Shvets, Alex and Kurmann, Thomas and Zhang, Zichen and Duggal, Rahul and Su, Yun-Hsuan and Rieke, Nicola and Laina, Iro and Kalavakonda, Niveditha and Bodenstedt, Sebastian and others},
	year         = 2019,
	journal      = {arXiv preprint arXiv:1902.06426}
}

@article{Allan2020EndoVis18,
	title        = {2018 robotic scene segmentation challenge},
	author={Allan, Max and Kondo, Satoshi and Bodenstedt, Sebastian and Leger, Stefan and Kadkhodamohammadi, Rahim and Luengo, Imanol and Fuentes, Felix and Flouty, Evangello and Mohammed, Ahmed and Pedersen, Marius and others},
	year         = 2020,
	journal      = {arXiv preprint arXiv:2001.11190}
}

@article{jin2022exploring,
	title        = {Exploring intra-and inter-video relation for surgical semantic scene segmentation},
	author       = {Jin, Yueming and Yu, Yang and Chen, Cheng and Zhao, Zixu and Heng, Pheng-Ann and Stoyanov, Danail},
	year         = 2022,
	journal      = {IEEE Trans. Med. Imag.},
	publisher    = {IEEE},
	volume       = 41,
	number       = 11,
	pages        = {2991--3002}
}

@inproceedings{ayobi2023matis,
	title        = {MATIS: Masked-attention transformers for surgical instrument segmentation},
	author       = {Ayobi, Nicol{\'a}s and P{\'e}rez-Rond{\'o}n, Alejandra and Rodr{\'i}guez, Santiago and Arbel{\'a}ez, Pablo},
	year         = 2023,
	booktitle    = {Proc. IEEE Int. Symp. Biomed. Imag.},
	pages        = {1--5}
}

@article{zadeh2023surgai3,
	title        = {SurgAI3. 8K: a labeled dataset of gynecologic organs in laparoscopy with application to automatic augmented reality surgical guidance},
	author={Zadeh, Sabrina Madad and Fran{\c{c}}ois, Tom and Comptour, Aur{\'e}lie and Canis, Michel and Bourdel, Nicolas and Bartoli, Adrien},
	year         = 2023,
	journal      = {J. Minim. Invasive Gynecol.},
	volume       = 30,
	number       = 5,
	pages        = {397--405}
}

@article{psychogyios2023sar,
	title        = {Sar-rarp50: Segmentation of surgical instrumentation and action recognition on robot-assisted radical prostatectomy challenge},
	author={Psychogyios, Dimitrios and Colleoni, Emanuele and Van Amsterdam, Beatrice and Li, Chih-Yang and Huang, Shu-Yu and Li, Yuchong and Jia, Fucang and Zou, Baosheng and Wang, Guotai and Liu, Yang and others},
	year         = 2023,
	journal      = {arXiv preprint arXiv:2401.00496}
}

@inproceedings{oh2019video,
	title        = {Video object segmentation using space-time memory networks},
	author       = {Oh, Seoung Wug and Lee, Joon-Young and Xu, Ning and Kim, Seon Joo},
	year         = 2019,
	booktitle    = ICCV,
	pages        = {9226--9235}
}

@article{yang2024samurai,
	title        = {Samurai: Adapting segment anything model for zero-shot visual tracking with motion-aware memory},
	author       = {Yang, Cheng-Yen and Huang, Hsiang-Wei and Chai, Wenhao and Jiang, Zhongyu and Hwang, Jenq-Neng},
	year         = 2024,
	journal      = {arXiv preprint arXiv:2411.11922}
}

@inproceedings{cheng2022xmem,
	title        = {Xmem: Long-term video object segmentation with an atkinson-shiffrin memory model},
	author       = {Cheng, Ho Kei and Schwing, Alexander G},
	year         = 2022,
	booktitle    = {Proc. Eur. Conf. Comput. Vis.},
	pages        = {640--658},
	organization = {Springer}
}

@misc{IntuitiveSurgical2024daVinci,
	title        = {{Da Vinci instruments}},
	author       = {{Intuitive Surgical, I.}},
	year         = 2024,
	note         = {Accessed: 27-Feb-2025},
	howpublished = {\url{https://www.intuitive.com/en-us/products-and-services/da-vinci/instruments}}
}

@inproceedings{resurgsam2,
  title={Resurgsam2: Referring segment anything in surgical video via credible long-term tracking},
  author={Liu, Haofeng and Gao, Mingqi and Luo, Xuxiao and Wang, Ziyue and Qin, Guanyi and Wu, Junde and Jin, Yueming},
  booktitle={Proc. Int. Conf. Med. Image Comput. Comput.-Assist. Intervent.},
  pages={435--445},
  year={2025},
  organization={Springer}
}

@inproceedings{li2023robust,
  title={Robust referring video object segmentation with cyclic structural consensus},
  author={Li, Xiang and Wang, Jinglu and Xu, Xiaohao and Li, Xiao and Raj, Bhiksha and Lu, Yan},
  booktitle=ICCV,
  pages={22236--22245},
  year={2023}
}

@article{wu2024toward,
  title={Toward robust referring image segmentation},
  author={Wu, Jianzong and Li, Xiangtai and Li, Xia and Ding, Henghui and Tong, Yunhai and Tao, Dacheng},
  journal={IEEE Trans. Image Process.},
  volume={33},
  pages={1782--1794},
  year={2024},
  publisher={IEEE}
}

@article{sam3,
  title={Sam 3: Segment anything with concepts},
  author={Carion, Nicolas and Gustafson, Laura and Hu, Yuan-Ting and Debnath, Shoubhik and Hu, Ronghang and Suris, Didac and Ryali, Chaitanya and Alwala, Kalyan Vasudev and Khedr, Haitham and Huang, Andrew and others},
  journal={arXiv preprint arXiv:2511.16719},
  year={2025}
}

@article{liu2025sam2s,
  title={SAM2S: Segment Anything in Surgical Videos via Semantic Long-term Tracking},
  author={Liu, Haofeng and Wang, Ziyue and Mishra, Sudhanshu and Gao, Mingqi and Qin, Guanyi and Low, Chang Han and Kong, Alex YW and Jin, Yueming},
  journal={arXiv preprint arXiv:2511.16618},
  year={2025}
}

@inproceedings{yin2025memory,
  title={Memory-augmented sam2 for training-free surgical video segmentation},
  author={Yin, Ming and Wang, Fu and Ye, Xujiong and Meng, Yanda and Fu, Zeyu},
  booktitle={Proc. Int. Conf. Med. Image Comput. Comput.-Assist. Intervent.},
  pages={328--337},
  year={2025},
}

@inproceedings{radford2023robust,
  title={Robust speech recognition via large-scale weak supervision},
  author={Radford, Alec and Kim, Jong Wook and Xu, Tao and Brockman, Greg and McLeavey, Christine and Sutskever, Ilya},
  booktitle={Proc. Int. Conf. Mach. Learn.},
  pages={28492--28518},
  year={2023},
}

@inproceedings{dam4sam,
	title        = {A distractor-aware memory for visual object tracking with sam2},
	author       = {Videnovic, Jovana and Lukezic, Alan and Kristan, Matej},
	year         = 2025,
	booktitle    = CVPR,
	pages        = {24255--24264}
}

@article{yang2025vivim,
  title={Vivim: a video vision mamba for ultrasound video segmentation},
  author={Yang, Yijun and Xing, Zhaohu and Yu, Lequan and Fu, Huazhu and Huang, Chunwang and Zhu, Lei},
  journal=CSVT,
  year={2025},
  publisher={IEEE}
}

@article{ma2025medsam2,
  title={Medsam2: Segment anything in 3d medical images and videos},
  author={Ma, Jun and Yang, Zongxin and Kim, Sumin and Chen, Bihui and Baharoon, Mohammed and Fallahpour, Adibvafa and Asakereh, Reza and Lyu, Hongwei and Wang, Bo},
  journal={arXiv preprint arXiv:2504.03600},
  year={2025}
}

@inproceedings{wei2026moves,
  title={Where It Moves, It Matters: Referring Surgical Instrument Segmentation via Motion},
  author={Wei, Meng and Yuan, Kun and Li, Shi and Zhou, Yue and Bai, Long and Navab, Nassir and Ren, Hongliang and Lee, Hong Joo and Vercauteren, Tom and Padoy, Nicolas},
  booktitle={Proc. AAAI Conf. Artif. Intell.},
  year={2026}
}

@article{madani2022artificial,
  title={Artificial intelligence for intraoperative guidance: using semantic segmentation to identify surgical anatomy during laparoscopic cholecystectomy},
  author={Madani, Amin and Namazi, Babak and Altieri, Maria S and Hashimoto, Daniel A and Rivera, Angela Maria and Pucher, Philip H and Navarrete-Welton, Allison and Sankaranarayanan, Ganesh and Brunt, L Michael and Okrainec, Allan and others},
  journal={Ann. Surg.},
  volume={276},
  number={2},
  pages={363--369},
  year={2022},
  publisher={LWW}
}

@article{ahmed2024deep,
  title={Deep learning for surgical instrument recognition and segmentation in robotic-assisted surgeries: a systematic review},
  author={Ahmed, Fatimaelzahraa Ali and Yousef, Mahmoud and Ahmed, Mariam Ali and Ali, Hasan Omar and Mahboob, Anns and Ali, Hazrat and Shah, Zubair and Aboumarzouk, Omar and Al Ansari, Abdulla and Balakrishnan, Shidin},
  journal={Artif. Intell. Rev.},
  volume={58},
  number={1},
  pages={1},
  year={2024},
  publisher={Springer}
}

@article{long2025surgical,
  title={Surgical embodied intelligence for generalized task autonomy in laparoscopic robot-assisted surgery},
  author={Long, Yonghao and Lin, Anran and Kwok, Derek Hang Chun and Zhang, Lin and Yang, Zhenya and Shi, Kejian and Song, Lei and Fu, Jiawei and Lin, Hongbin and Wei, Wang and others},
  journal={Sci. Robot.},
  volume={10},
  number={104},
  pages={eadt3093},
  year={2025},
  publisher={American Association for the Advancement of Science}
}

@inproceedings{sam2,
    title={{SAM} 2: Segment Anything in Images and Videos},
    author={Ravi, Nikhila and Gabeur, Valentin and Hu, Yuan-Ting and Hu, Ronghang and Ryali, Chaitanya and Ma, Tengyu and Khedr, Haitham and R{\"a}dle, Roman and Rolland, Chloe and Gustafson, Laura and others},
    booktitle=ICLR,
    year={2025}
}

@article{zhou2022survey,
  title={A survey on deep learning technique for video segmentation},
  author={Zhou, Tianfei and Porikli, Fatih and Crandall, David J and Van Gool, Luc and Wang, Wenguan},
  journal=PAMI,
  volume={45},
  number={6},
  pages={7099--7122},
  year={2022},
  publisher={IEEE}
}

@article{wang2025lacoste,
  title={LACOSTE: Exploiting stereo and temporal contexts for surgical instrument segmentation},
  author={Wang, Qiyuan and Zhao, Shang and Xu, Zikang and Zhou, S Kevin},
  journal={Med. Image Anal.},
  volume={99},
  pages={103387},
  year={2025},
  publisher={Elsevier}
}

@article{ceron2022real,
  title={Real-time instance segmentation of surgical instruments using attention and multi-scale feature fusion},
  author={Cer{\'o}n, Juan Carlos {\'A}ngeles and Ruiz, Gilberto Ochoa and Chang, Leonardo and Ali, Sharib},
  journal={Med. Image Anal.},
  volume={81},
  pages={102569},
  year={2022},
  publisher={Elsevier}
}

@article{chen2024ma,
  title={Ma-sam: Modality-agnostic sam adaptation for 3d medical image segmentation},
  author={Chen, Cheng and Miao, Juzheng and Wu, Dufan and Zhong, Aoxiao and Yan, Zhiling and Kim, Sekeun and Hu, Jiang and Liu, Zhengliang and Sun, Lichao and Li, Xiang and others},
  journal={Med. Image Anal.},
  volume={98},
  pages={103310},
  year={2024},
  publisher={Elsevier}
}

@article{zhao2025rethinking,
  title={Rethinking data imbalance in class incremental surgical instrument segmentation},
  author={Zhao, Shifang and Bai, Long and Yuan, Kun and Li, Feng and Yu, Jieming and Dong, Wenzhen and Wang, Guankun and Hoque, Mobarak Islam and Padoy, Nicolas and Navab, Nassir and others},
  journal={Med. Image Anal.},
  volume={105},
  pages={103728},
  year={2025},
  publisher={Elsevier}
}

@article{wu2025medical,
  title={Medical sam adapter: Adapting segment anything model for medical image segmentation},
  author={Wu, Junde and Wang, Ziyue and Hong, Mingxuan and Ji, Wei and Fu, Huazhu and Xu, Yanwu and Xu, Min and Jin, Yueming},
  journal={Med. Image Anal.},
  volume={102},
  pages={103547},
  year={2025},
  publisher={Elsevier}
}

@article{vandenhende2021multi,
  title={Multi-task learning for dense prediction tasks: A survey},
  author={Vandenhende, Simon and Georgoulis, Stamatios and Van Gansbeke, Wouter and Proesmans, Marc and Dai, Dengxin and Van Gool, Luc},
  journal=PAMI,
  volume={44},
  number={7},
  pages={3614--3633},
  year={2021},
  publisher={IEEE}
}

@article{schmidgall2025will,
  title={Will your next surgeon be a robot? Autonomy and AI in robotic surgery},
  author={Schmidgall, Samuel and Opfermann, Justin D and Kim, Ji Woong and Krieger, Axel},
  journal={Sci. Robot.},
  volume={10},
  number={104},
  pages={eadt0187},
  year={2025},
  publisher={American Association for the Advancement of Science}
}

@article{zou2023segment,
  title={Segment everything everywhere all at once},
  author={Zou, Xueyan and Yang, Jianwei and Zhang, Hao and Li, Feng and Li, Linjie and Wang, Jianfeng and Wang, Lijuan and Gao, Jianfeng and Lee, Yong Jae},
  journal=NeurIPS,
  volume={36},
  pages={19769--19782},
  year={2023}
}

@inproceedings{pei2026synergistic,
  title={Synergistic Bleeding Region and Point Detection in Laparoscopic Surgical Videos},
  author={Pei, Jialun and Zhou, Zhangjun and Guo, Diandian and Li, Zhixi and Qin, Jing and Du, Bo and Heng, Pheng-Ann},
  booktitle=CVPR,
  year={2026}
}

@article{pei2025instrument,
  title={Instrument-tissue-guided surgical action triplet detection via textual-temporal trail exploration},
  author={Pei, Jialun and Zhang, Jiaan and Qin, Guanyi and Wang, Kai and Jin, Yueming and Heng, Pheng-Ann},
  journal={IEEE Trans. Image Process.},
  year={2025},
  publisher={IEEE}
}

\end{document}